\documentclass{ptephy_v1}%%%%%% to generate preprint number with ptep logo

\newcommand{\xbar}[1]{#1 \hspace{-5.5pt}/}
\begin{document}

\title{Origin of Mass - Horizons Expanding from the Nambu's Theory
%\footnote{Talk given at Yoichiro Nambu Memorial Symposium, Sept. 29, 2015, Osaka City University.}
}

%%%% To generate auto affiliation numbers please use \author{}\affil{} command

\author{Koichi Yamawaki}
\affil{Kobayashi-Maskawa Institute for the Origin of Particles and the Universe, Nagoya University, Nagoya 464-8602, Japan %\email{yamawaki@kmi.nagoya-u.ac.jp}
}

%\author{Insert second author name here}
%\affil{Insert second author address here}

%\author{Insert third author name here}
%\author[3]{Insert fourth author name here} %%% Use optional bracket [3] to change the respective address
%\affil{Insert third author address here}

%\author{Insert last author name here\thanks{These authors contributed equally to this work}}
%\affil{Insert last author address here}

%%% To include the collaborator name... Please use the command "\collaborator"
%%% For example: \collaborator{ATLAS Collaboration}

\begin{abstract}%
The visible (Matter)  has no difference from the invisible (Vacuum), the invisible has no difference from the visible.
 The visible is nothing but the invisible,  the invisible is nothing but the visible. 
  
  - Heart Sutra (translation by KY)\\

Origin of mass may be strong dynamics of matter in the vacuum. 
Since the initial proposal of Nambu for the origin of the nucleon mass, the dynamical symmetry breaking in the {\it strongly  coupled underlying theories} 
has been expanding
the  horizons in the context of the modern version of the origin of mass 
beyond the Standard Model (SM).

The Nambu-Jona-Lasinio (NJL) model is a typical strong coupling theory with 
the  {\it non-zero critical coupling and a  large anomalous dimension} $\gamma_m=2$,  in sharp contrast to its precedent model, the Bardeen-Cooper-Schrieffer  theory for the superconductor.
 The non-zero critical coupling is also hidden in the asymptotically free gauge theories including QCD and walking technicolor: it reveals itself in the chiral symmetry restoration  
where the coupling cannot grow above the ``hidden'' critical coupling in the infrared region (infrared conformality).

 As is well known, the NJL model can be cast into the SM Higgs Lagrangian.
We show 
that the {\it SM Higgs Lagrangian  
is simply rewritten into a form of 
the (approximately) scale-invariant nonlinear sigma model}, with both the chiral symmetry and scale symmetry realized nonlinearly, with   
the SM Higgs being nothing but the (pseudo-) dilaton.
The SM Higgs Lagrangian is {\it further gauge equivalent to the scale-invariant Hidden Local Symmetry (HLS) Lagrangian}, s-HLS, having spin 1 bosons hidden in the SM.
 
As the simplest possible underlying theory for the SM Higgs Lagrangian we first discuss  the top quark condensate (``top-mode SM'') based on the (scale-invariant) NJL model
with only top (plus possibly bottom) coupling larger than the critical coupling, where the top-mode dilaton is the 125 GeV Higgs and the HLS gauge boson (``top-mode rho meson'') (and the top-mode axion)
may be detected at LHC.

We then discuss 
the walking technicolor having near infrared conformality 
and large anomalous dimension $\gamma_m\simeq 1$. Its effective theory is the s-HLS model precisely the same as the SM Higgs Lagrangian (with larger chiral symmetry), where the 125 GeV Higgs is successfully identified with the techidilaton. The 2 TeV diboson and 750 GeV diphoton excesses at LHC are identified with the HLS technirho and the technipion, respectively. 
\end{abstract}

\maketitle

\section{Introduction}

Professor Nambu made great achievements in so much inexhaustible depth and wideness, and thus it may be something like the picture of ``The Blind Men and the Elephant'' to talk
about only a single aspect of his physics. But the subject I am going to talk about is not just one of them, but probably his most influential one. In fact the 2018 Nobel Prize announcement is
``for the discovery of the mechanism of spontaneous broken symmetry in subatomic physics'' \cite{2008Nobel}.
His spontaneous symmetry breaking (SSB)  \cite{Nambu:1961tp} was the theory for the origin of mass of  the nucleon, then an elementary particle, generated dynamically from nothing (the vacuum) through 
the nucleon-antinucleon pair condensate. Although the nucleon is no longer an elementary particle, the essence of his mechanism to generate mass of composite particles, hadrons including the nucleon, as well as the near masslessness of the 
composite pion, 
 is now realized through the quark-antiquark 
condensate in the underlying theory QCD. This mass constitutes 99 \% of mass of the nucleon, namely of the ordinary matter made out of the atoms, and thus the Nambu's theory already accounted for the
origin of the dominant part of the mass of the visible world. 

The problem of the origin of mass in the modern particle physics is only for the rest 1\% of the mass of the matter, the elementary particles
of the Standard Model (SM), which is attributed to 
the Higgs boson whose origin is still mysterious. I will discuss that this 1\% may also be explained by the dynamical symmetry breaking in some underlying theory, similarly to the Nambu's theory.

Origin of Mass of all the SM particles is the Higgs VEV $v=\sqrt{\frac{-\mu_0^2}{\lambda}}=246\,{\rm GeV}$ or the Higgs mass $M_\phi^2 =2\lambda v^2=-2\mu_0^2$ read from  the SM Higgs Lagrangian:
\begin{eqnarray}
{\cal L}_{\rm Higgs}&=&
 |\partial_\mu h|^2 -\mu_0^2 |h|^2 -\lambda|h|^4
 \label{Higgs}\\
&=& \frac{1}{2} \left[
\left(\partial_\mu {\hat \sigma}\right)^2 +\left(\partial_\mu {\hat \pi_a}\right)^2
\right]
-\frac{1}{2}
\mu_0^2  \left[ 
{\hat \sigma}^2+{\hat \pi_a}^2 
\right]-\frac{\lambda}{4} \left[ 
{\hat \sigma}^2+{\hat \pi_a}^2 
\right]^2 
 \label{sigma}\\ 
&=& \frac{1}{2} {\rm tr} \left( \partial_\mu M\partial^\mu M^\dagger \right)
 - \left[\frac{\mu_0^2}{2} {\rm tr}\left(M M^\dagger\right)+\frac{\lambda}{4}  \left({\rm tr}\left(M M^\dagger\right)\right)^2\right] \,,
 \label{sigmaMatrix}
 \end{eqnarray}
where we have rewritten the conventional form in Eq.(\ref{Higgs})  into the Gell-Mann-Levy (GL) $SU(2)_L \times SU(2)_R$ linear sigma model \cite{GellMann:1960np} in Eq.(\ref{sigma})  through
\begin{equation}
h=\left(\begin{array}{c}
\phi^+\\
\phi^0\end{array}  \right)=\frac{1}{\sqrt{2}} \left(\begin{array}{c}
i{\hat \pi}_1+{\hat \pi}_2\\
\hat{\sigma}-i {\hat \pi}_3\end{array}\right)\,,
\end{equation} 
and further into Eq.(\ref{sigmaMatrix}) with the $2\times 2$ matrix $M$ 
\begin{equation}
M=(i \tau_2 h^*, h) = \frac{1}{\sqrt{2}}\left({\hat \sigma}\cdot  1_{2\times 2} +2i {\hat \pi}\right)\, \quad \left({\hat \pi} \equiv {\hat \pi}_a \frac{\tau_a}{2}\right)\,,
\end{equation}
which transforms under $G=SU(2)_L\times SU(2)_R$ as:
\begin{equation}
M \rightarrow g_L \, M\, g_R^\dagger \,,\quad \left(g_{R,L} \in SU(2)_{R,L}\right)\,.
\end{equation}
Then the origin of mass 
is attributed to  the mysterious input mass parameter of the {\it tachyons}, $\hat \pi$ and $\hat \sigma$ (not physical particles), with the mass $\mu_0$ such that 
\begin{equation}
\mu_0^2<0
\end{equation}
 {\it as a free parameter}.
But why the tachyon? How is the tachyon mass determined?  
SM cannot answer to these questions, even though the Higgs boson has been discovered with the mass near 125 GeV. 

Historically, the GL linear sigma model  in the form of Eq.(\ref{sigma}) as the prototype of the Higgs Lagrangian Eq.(\ref{Higgs}) was proposed for phenomenologically describing the pion (as well as the nucleon) 
{\it without concept of the SSB}, while the Nambu-Jona-Lasinio (NJL) model \cite{Nambu:1961tp} explained the same property at the deeper level in terms of the dynamical symmetry breaking
due to the vacuum property. 
Here we should
 recall that 
{\it the SSB was born as a dynamical symmetry breaking (DSB)}, where the  tachyons are in fact generated as  composites of the dynamical consequence of the strong dynamics, but not  ad hoc inputs as in the GL theory. Actually, the GL theory is now regarded as an effective theory (macroscopic theory) for the NJL model as a microscopic theory, as is the Ginzburg-Landau (GL) theory
for the Bardeen-Cooper-Schrieffer (BCS) theory \cite{BCS} for the superconductor. So history may repeat itself. 

I first discuss that although his idea was motivated by the BCS theory, it was not just a copy of it but essentially new in the most important
aspect, namely it created a new dynamics, although based on the same kind of four-fermion interaction: the NJL dynamics is the {\it strong coupling theory having non-zero critical coupling} to separate the SSB phase with $\mu_0^2<0$  (above the critical coupling) from the non-SSB phase with $\mu_0^2>0$ (below the critical coupling). It is in sharp contrast to the BCS theory which is a {\it weak coupling theory having the zero critical coupling}, always in the SSB phase $\mu_0^2<0$  even for infinitesimal (attractive) coupling,  
due to the Fermi surface. The Fermi surface reduces the effective dimensions by 2 so as to make the theory in effectively $1+1$ dimensions like the Thirring model and/or Gross-Neveu model.
 
The non-zero critical coupling is also hidden in the asymptotically free gauge theories including the QCD:  it reveals itself in the chiral symmetry restoration  
 when the system is in the extreme condition such as the
 high temperature, high density, and large number of light fermions, where the coupling cannot grow above the hidden critical coupling in the infrared region as strong enough  to form  the fermion-antifermion
 condensate.
The existence of the non-zero critical coupling in the gauge theory was first recognized by Maskawa-Nakajima \cite{Maskawa:1974vs} in the ladder Schwinger-Dyson (SD) equation, a gauge theory analogue of the NJL gap equation.  The solution of the ladder SD equation in the weak coupling region (existing even for the infinitesimal coupling) \cite{Johnson:1964da} disappears at zero fermion bare mass at finite cutoff,
which is actually the explicit chiral symmetry breaking solution vanishing when the cutoff is removed. \footnote{
At TV interview after announcement of the Nobel prize together with Professor Nambu in 2008, Toshihide Maskawa confessed that the paper he most studied was the Nambu's paper on SSB,
``I exhausted it''. He 
in fact discovered the non-zero critical coupling  for SSB in the gauge theory \cite{Maskawa:1974vs} not just in the NJL four-fermion model. 
At that time I was a graduate student at Kyoto University to hear it first hand and have been influenced by this  work, strong coupling gauge theory (SCGT) with non-zero criticality,  ever since. See Nagoya SCGT workshops, http://www.kmi.nagoya-u.ac.jp/workshop/SCGT15/ }   
Although the asymptotically-free gauge theory like QCD has no explicit critical coupling to divide the SSB phase from non-SSB phase (having only a single phase of SSB), the running coupling always becomes strong in the infrared region where the coupling exceeds a hidden critical coupling to trigger the SSB having the condensate of order of the scale of this mass region \cite{Higashijima:1983gx}.

The main purpose of this article is to describe the expanding horizon of such a strong coupling dynamics characterized by the non-zero critical coupling initiated by Professor Nambu in view of the
modern version of  the
origin of mass,
 namely the composite Higgs models having large anomalous dimension. First,  the (weakly gauged) strong coupling four-fermion models like the top quark condensate model \cite{Miransky:1988xi,Nambu:1989jt,Bardeen:1989ds}, where only
the top quark has the strong coupling  above the criticality (anomalous dimension $\gamma_m \simeq 2$ \cite{Miransky:1988gk}) so as to be responsible for the electroweak symmetry breaking \cite{Miransky:1988xi}. Second, 
the gauge theories such as the walking technicolor based on the
near conformal gauge theory just above the criticality having anomalous dimension $\gamma_m\simeq1$ and a 
composite dilaton (technidilaton) as the composite Higgs \cite{Yamawaki:1985zg,Bando:1986bg}.  
The technidilaton in the walking technicolor has been shown to be consistent with the 125 GeV Higgs at the present LHC experimental data \cite{Matsuzaki:2012mk,Matsuzaki:2012xx,   Matsuzaki:2015sya}.   

Before discussing possible underlying theory for SM, I show that the SM Higgs Lagrangian itself already has some hints  for the
theory beyond the SM. It was shown \cite{Fukano:2015zua} that the SM Higgs Lagrangian itself possesses nonlinearly realized ``hidden'' symmetries (scale symmetry and Hidden Local Symmetry (HLS) \cite{Bando:1984ej,Bando:1987br, Harada:2003jx}, both spontaneously broken),  in addition to the well-known symmetry, nonlinearly realized global $SU(2)_L\times SU(2)_R$  chiral symmetry (also spontaneously broken)
to be gauged by the electroweak symmetry. It is in fact straightforward to show \cite{Fukano:2015zua} that the SM Higgs Lagrangian is cast into the scale-invariant nonlinear chiral Lagrangian \cite{Matsuzaki:2012mk}, 
and then 
further shown to be
gauge equivalent to the scale-invariant HLS (s-HLS) Lagrangian \cite{Kurachi:2014qma}. The SM Higgs is nothing but a (pseudo-) dilaton! \cite{Fukano:2015zua}
{\it This is the very nature of the SM Higgs Lagrangian, quite independent of details of the
possible underlying theory as the UV completion.} 
Also the HLS  can
naturally accommodate the vector bosons, analogues of the rho mesons in the QCD, into the SM (``SM rho meson'') \cite{MY16}: 
It would be  the simplest extension of the SM to account for the 2 TeV diboson events at LHC \cite{Aad:2015owa}.

Then I elaborate \cite{Yamawaki:2015tmu} on the well-known fact that the NJL model %, a four-fermion theory,  
can be regarded as the microscopic theory (underlying theory or ultraviolet (UV) completion) for the SM Higgs Lagrangian,
or the GL linear sigma model,  
as the macroscopic theory (effective theory) at composite level. With the coupling larger than the non-zero critical coupling, the NJL model equivalent to the SM Higgs has also the nonlinearly realized hidden (approximate) scale symmetry for the SM Higgs as a composite pseudo-dilaton (``NJL dilaton''), together with  the HLS for the dormant composite spin 1 boson (``NJL rho meson'') as a possible
candidate for the LHC diboson events \cite{Aad:2015owa}.
Although both are trivial theories having no interaction in the infinite cutoff limit (Gaussian fixed point), I will discuss possible way out, one \cite{Kondo:1991yk} being the gauged NJL model
in combination with the walking gauge theory,   another \cite{Yamawaki:2015tmu} the recently suggested  different way of the continuum limit where the composite Higgs becomes massless
(up to the trace anomaly) as the pseudo-dilaton in the same sense as the SM Higgs.

The simplest possibility for such a composite model would be the top quark condensate model  (``top-mode SM'') \cite{Miransky:1988xi,Nambu:1989jt,Bardeen:1989ds}, where  crucial is the
non-zero criticality \cite{Miransky:1988xi}: only top (may also bottom) has the coupling larger than  the non-zero critical coupling to acquire the dynamical mass due to SSB.
Near the scale-invariant limit, the top-mode dilaton may be  the 125 GeV Higgs, and the HLS gauge boson (``top-mode rho meson'') may be identified with the recent 2 TeV diboson excess (and the top-mode axion, $\bar b b$ bound state, may be identified with the 750 GeV diphoton excess at LHC \cite{750} which was reported after this symposium).

We then discuss the walking technicolor proposed based on the SSB solution of the ladder SD equation to have a large anomalous dimension $\gamma_m=1$ and technidilaton as a pseudo-Nambu-Goldstone (NG) boson of the approximate scale symmetry \cite{Yamawaki:1985zg,Bando:1986bg}. Such a scale-symmetric walking gauge theory may be realized when flavor number $N_F$ of massless fermions  
is large in the asymptotically free gauge theory (``large $N_F$ QCD'')  \cite{Appelquist:1996dq,Miransky:1996pd}, with $N_F (\gg 2)$ slightly smaller than that having an infrared fixed point  (conformal window)  where the coupling in the infrared region is almost constant and below the critical coupling 
so that the SSB does not take place. The effective theory of the walking technicolor is the s-HLS Lagrangian with a larger chiral symmetry $SU(N_F)_L \times SU(N_F)_R$, with typically $N_F=8$ (one-family model), precisely the same type of the s-HLS as in the case for the SM Higgs Lagrangian with $SU(2)_L \times SU(2)_R$. The technidilaton as a composite Higgs has been shown  \cite{Matsuzaki:2012mk,Matsuzaki:2012xx, Matsuzaki:2015sya} to be
consistent with the
present LHC 125 GeV Higgs, and the HLS vector mesons (walking technirhos) have also been shown \cite{Fukano:2015hga} to be consistent with the LHC diboson events \cite{Aad:2015owa}.
 (We also showed \cite{Matsuzaki:2015che}
 that one of the technipions can be identified consistently with the 750 GeV diphoton events at LHC \cite{750} reported after the symposium).   

Several theoretical issues are discussed such as the recent lattice studies of the walking theories, as well as the ladder,  and  the renormalizability of the gauged-NJL model and the conformal phase transition, etc.

\section{NJL the Strong Dynamics vs. BCS the Weak Dynamics}

It is widely believed that the NJL model is a copy of the BCS. Here I emphasize that they are essentially different dynamics, 
NJL as the strong coupling with critical coupling no-zero, while the BCS as a weak coupling with the critical coupling zero.
The difference comes from the Fermi surface in BCS which reduces the effective phase space from 3+1 to 1+1, while the NJL case is in the free space
of  full 3+1 dimensions. The attractive forces are more efficient in smaller phase space.

 Let us start with the $SU(2)_L \times SU(2)_R$ NJL model \cite{Nambu:1961tp}  for $N_C$  2-flavored Dirac fermions $\psi$:
\begin{equation}
{\cal L}_{\rm NJL} = \bar \psi i\gamma^\mu\partial_\mu \psi + \frac{G}{2} \left[ (\bar \psi \psi)^2 +(\bar \psi i \gamma_5 \tau^a \psi)^2\right]\,.
\end{equation}
When the fermion-antifermion condensate in the vacuum takes place, $\langle 0|\bar \psi \psi|0\rangle \ne0$, it reads 
\begin{equation}
{\cal L}_{\rm NJL} -\bar \psi i\gamma^\mu\partial_\mu \psi = \ G  \langle \bar \psi \psi\rangle\, \bar \psi \psi +\cdots=
-m_F \, \bar \psi \psi +\cdots\,.
\end{equation}
 At the $1/N_C$ leading order this yields the self-consistent 
NJL gap equation for the dynamical mass $m_F$  of $\psi$:
\begin{equation}
m_F= -G \langle \bar \psi \psi\rangle=G\, {\rm Tr}[
 S_F(p)] =G \cdot 4 N_C \int \frac{d^4p}{i (2\pi)^4} \frac{m_F}{m_F^2-p^2}
 \label{NJLgap}
  \end{equation}
which has an SSB solution $m_F\ne 0$:
\begin{equation}
\frac{1}{G}-\frac{1}{G_{\rm cr}} =\frac{\Lambda^2}{4\pi^2} \left(\frac{1}{g} -\frac{1}{g_{\rm cr}}\right)= - \frac{1}{4\pi^2} N_C m_F^2 \ln \left(\frac{\Lambda^2}{m_F^2}
\right) <0 \,,
\label{SSBNJL}
\end{equation}
only for the strong coupling 
\begin{equation}
G>G_{\rm cr}=\frac{4\pi^2}{N_C \Lambda^2}\ne 0
\quad
 (g\equiv\frac{G\Lambda^2}{4\pi^2} >g_{\rm cr}=\frac{1}{N_c}\ne 0)\,.
 \label{strong}
 \end{equation}

We shall later discuss that this in fact corresponds to the tachyon mass $\mu_0^2<0$ in Eq.(\ref{sigma}): $\mu_0^2 = (\frac{1}{G} - \frac{1}{G_{\rm cr}})\cdot Z_\phi^{-1}= - 2m_F^2<0$, where
$Z_\phi=\frac{N_C}{8\pi^2} \ln \frac{\Lambda^2}{m_F^2}$. The (composite) tachyon has been induced dynamically by the $-1/G_{\rm cr}$ term due to the loop effects in the large $N_C$ limit. Of course the tachyon is not a physical particle, which simply implies instability of the trivial vacuum with $m_F=0$
(no SSB).
For the weak coupling $G<G_{\rm cr}$ there exists only the non SSB solution $m_F\equiv 0$ where no tachyon exists. 

Note that ``strong coupling'' as defined by non-zero critical coupling
does not necessarily mean
numerically strong, particularly in the large $N_C$ limit $g_{\rm cr} = 1/N_C \ll 1$. However, the attractive forces  in the condensate channel are not from a single fermion but actually from sum of all the $N_c$ fermions coherently, which ends up with  really strong  $N_C g_{\rm cr} =  {\cal O}(1)$. [As we discuss later,
this also applies to the strong coupling gauge theory where $N_C \alpha_{\rm cr} ={\cal O} (1)$, while the gauge coupling criticality itself $\alpha_{\rm cr} \sim 1/N_C \ll 1$
is negligibly small (but non-zero)  in the large $N_C$ limit].
 
In contrast to the non-zero critical coupling of the NJL model, the BCS theory for the superconductor has the zero critical coupling (``weak coupling theory'') 
due to the electron Fermi surface $E_F= \frac{p_F^2}{2m_e}$ ($m_e$: electron mass in the free space), which affects the fermion-fermion condensate $\langle \psi \psi\rangle\ne 0$ (dynamical Majorana mass $\Delta$ as the gap) instead of fermion-antifermion condensate $\langle \bar \psi \psi\rangle\ne 0 $ . The essence can be read from the effective dimension of the momentum in the integral of the gap equation of the BCS $\int \frac{d^4 p}{i (2\pi)^4}=\int \frac{d^3 p}{(2\pi)^3} \frac{d \omega}{i (2\pi)}$:
\begin{equation}
 \int \frac{d^3 p}{(2\pi)^3}
 =\int \frac{(4\pi p^2) d p}{(2\pi)^3} 
\Rightarrow  
4\pi p_F^2  \int_{|E(p)-E_F|<\omega/2} \frac{d p}{(2\pi)^3}  = \frac{N}{2} \int_{E_F-\omega_D/2}^{E_F+\omega_D/2}  d E(p) \,,
\end{equation}
where  $E(p)\equiv \frac{p^2}{2m_e}$ and $N\equiv \frac{m_e p_F}{\pi^2}=$ constant:
The $3-$dimensional electron momentum $\overrightarrow{p}$ is confined to a one-dimensional direction normal to the Fermi surface $E_F=\frac{p_F^2}{2m_e}$ in the 
narrow energy shell bounded by the Debye energy $\omega_D$ (cutoff). After integral $\int d \omega$, with the fermion propagator $S_F(p)^{(Majorana)}={\cal F.F.}  \langle T(\psi(x) \psi(0)\rangle
= \Delta/(\omega^2 -|\Delta|^2 -E(p)^2)$ (
instead of ${\cal F.F.}  \langle T(\bar \psi(x) \psi(0)\rangle$) ),  the BCS gap equation 
corresponding to Eq.(\ref{NJLgap})  (Majorana mass without
factor 4) reads 
\begin{equation}
|\Delta| = G\frac{N}{2} \int_0^{E_F+\omega_D/2} d E(p) \frac{|\Delta| } {\sqrt{|\Delta|^2+E(p)^2}}\sim |\Delta| \left[ \frac{N G}{2} \ln\frac{|\Delta|}{\omega_D}
\right]\,.
\label{gapBCS}
\end{equation}
Then the SSB solution with $|\Delta|\ne 0$ exists even for infinitesimal coupling $1\gg N G >N G_{\rm crit}=0$:
\begin{equation}
|\Delta| \sim \omega_D \exp\left(-\frac{2}{N G}\right)\,,\quad \left(1\gg N G >N G_{\rm cr}=0\right)\,,
\label{BCSgap}
\end{equation}
in contrast to the SSB solution in NJL model in Eq.(\ref{SSBNJL}) with $N_C g>N_C g_{\rm cr} ={\cal O} (1) $ in Eq.(\ref{strong}).

 The result is intuitively obvious: The fermion pair in the one dimensional space is ``bound'' even for infinitesimal coupling, since there is no way
 to escape from each other, while that in the higher dimensional space can freely move from each other and hence needs strong attractive forces
 to bind it together. This is the effective dimensional reduction. 
The situation that the lower dimensional theory lowers the critical coupling can be viewed explicitly by the $D (1+1<D<3+1)$ dimensional four-fermion theory, the Gross-Neveu model, with
$D$ changed continuously \cite{Kikukawa:1989fw}. The gap equation is simply changed as $\int \frac{d^4p}{i(2\pi)^4} \Rightarrow \int \frac{d^D p}{i (2\pi)^D}$ in Eq. (\ref{NJLgap}). Similarly to Eq.(\ref{SSBNJL}), the SSB solution exists:\cite{Kikukawa:1989fw} 
 \begin{equation}
\frac{1}{g} -\frac{1}{g_{\rm cr}} 
=- \frac{N_C\xi_D}{2- \frac{D}{2}}  \cdot \left(\frac{m_F}{\Lambda}\right)^{D-2} <0 
 \,,
\label{SSBDNJL}
\end{equation} 
only  for the strong coupling;\footnote{
In $D>4$ dimensions, the form $g_{\rm cr}= \frac{1}{N_C} \left(\frac{D}{2}-1\right) $ remains the same, in accord with the above intuitive picture for the 
required binding force strength  depending on the phase volume,
while the gap equation takes a similar but different form:  
$ 1/g -1/g_{\rm cr} 
=- \frac{N_C}{D/2-2}  \cdot \left(\frac{m_F}{\Lambda}\right)^2$ \cite{Hashimoto:2000uk}\,.
}
\begin{equation}
g \,\, > \,\,  g_{\rm cr}= \frac{1}{N_C} \left(\frac{D}{2}-1\right) \rightarrow 0 \quad (D\rightarrow 2),
 \end{equation}
where $g \equiv G\Lambda^{D-2} \, \frac{2^{D/2}}{(4\pi)^{D/2} \Gamma(D/2)}$ and 
$\xi_D= B(\frac{D}{2}-1,3-\frac{D}{2}) \rightarrow \frac{1}{D/2-1}=\frac{1}{N_C g_{\rm cr}}$ for  $D\rightarrow 2$ ($\rightarrow 1 $ for $D\rightarrow 4$) \footnote{
If we take the $D\rightarrow 4$ limit, on the other hand, the gap equation is reduced to Eq.(\ref{SSBNJL}) except for the logarithmic factor. This log factor is a crucial difference between the $2< D<4$ and the $D=4$ four-fermion theories. As we discuss later, the former is renormalizable in $1/N_C$ expansion having the nontrivial fixed point at $g=g_{\rm cr}$  in the beta function, $\beta(g) =- \frac{D-2}{g_{\rm cr}} \, g\,(g-g_{\rm cr})$ \cite{Kikukawa:1989fw}, while the latter is not, a trivial theory, with the beta function having the Gaussian fixed point at $g=g_{\rm cr}$. ($g=0$ is an infrared fixed point defining the infrared free theory, with $g<0$ being the repulsive
forces.)  
}.
The critical coupling $g_{\rm cr}$ indeed decreases as $D$ does to vanish at $D=2$.
This yields for $D\rightarrow2$ the well-known result:
\begin{equation}
m_F=\Lambda \exp \left(-\frac{1}{2N_C g} \right)\,,\quad \left(N_C g > N_C g_{\rm cr}=0\right)\,,
\label{D2NJL}
\end{equation}
which is of the same form as Eq.(\ref{BCSgap}).

Thus the BCS dynamics in some sense is similar to the $D=1+1$ four-fermion theories
 such as the Thirring model and the Gross-Neveu model. 
 There is a caveat \cite{Witten:1978qu}, however: the genuine $D=1+1$ dimensional theory is not actually in the SSB phase in accord with the Merwin-Wagner-Coleman theorem, although it has a massless bound state and a massive
 fermion with mass of the form of Eq.(\ref{D2NJL}) in the large $N_C$ limit, similarly to the SSB phase. However, the absence of the NG boson and lack of SSB does not apply to the BCS theory in contrast to
 the Thirring model and Gross-Neveu model, since the BCS theory
 is not a genuine $1+1$ dimensional model but rather a brane model: only fermions (not anti-fermions) are confined to the $1+1$-brane, the Fermi surface, a consequence of the Fermi statistics, while 
 the fermion-fermion pair composite NG boson as a boson lives  freely from  the Fermi surface   in the full $3+1$ dimensional bulk,
 and hence SSB and NG boson do exist, in accord with the superfluidity and superconductor.
   
 To summarize the Nambu's approach to the origin of mass, 
 the theory having intrinsic mass scale $\Lambda\sim G^{-1/2}$ may or may not produce the particle mass $m_F$, depending on the coupling strength:
 the strong coupling dynamics for $G>G_{\rm cr}\ne 0$  creates the composite tachyon
 with negative ${\rm mass}^2$ $\mu_0^2 \sim 1/G -1/G_{\rm cr} = -\frac{N_C}{4\pi^2} m_F^2 \ln \frac{\Lambda^2}{m_F^2} <0$ in such a way that the particle mass  $m_F$ is generated from the intrinsic mass scale 
 $\Lambda$. By
 fine tuning the strong coupling $G (>G_{\rm cr}\ne 0)$ as $G\simeq G_{\rm cr}$, we can arrange a big hierarchy $m_F\ll \Lambda$ (near chiral symmetry restoration). 
 On the other hand, for the weak coupling $G<G_{\rm cr}$ there
 exists no particle mass $m_F\equiv 0$, although the theory has an intrinsic mass scale $\Lambda$. 
 This is   an essential difference from the BCS theory
 which has a zero critical coupling,  producing always a non-zero gap $\Delta\ne 0$ even for the infinitesimal coupling.

 As discussed later, this non-BCS phase structure of the NJL dynamics
 was in fact the original motivation of 
 the top quark condensate model of Ref. \cite{Miransky:1988xi}, where the top
 quark  having a coupling larger than the critical coupling is discriminated from others having those smaller than the critical coupling, so that 
  only the top has mass of order of weak scale in such a way as to produce only three NG bosons 
 responsible for the electroweak symmetry breaking.
 This is in contrast to the ``bootstrap symmetry breaking'' \cite{Nambu:1989jt} which is based on the BCS dynamics without the notion of the 
 non-zero criticality.
  
 \section{Strong Coupling Gauge Theories for the Origin of Mass} 
 \label{SCGT}
 
 The dynamical mass of the fermion $m_F$ picks up the intrinsic scale $\Lambda$ (cutoff) which regularizes the theory and brings  the explicit breaking of the scale symmetry
corresponding to the trace anomaly  in the renormalized quantum theory.  In the asymptotically free gauge theory $\Lambda$ can be identified with
the renormalization-group invariant intrinsic scale such as $\Lambda_{\rm QCD}$ induced by the perturbative trace anomaly, as we discuss later. 
 
There also exists a non-zero critical coupling for SSB in the gauge theory with massless fermion, 
as first noted \cite{Maskawa:1974vs} in the ladder SD equation,
 with non-running coupling $\alpha(\mu^2)\equiv \alpha=g^2/(4\pi)$ in the Landau gauge,
a straightforward extension of the NJL gap equation Eq. (\ref{NJLgap}), this time for the fermion mass function $\Sigma(-p^2)$ instead of the constant mass $m_F$
(For details see e.g., Ref.\cite{Matsuzaki:2015sya}): 
 \begin{equation}
  S_F^{-1}(p) \ =\ S^{-1}(p) + \int \frac{d^4 k}{(2 \pi)^4}\ 
  C_2\, 
  g^2 
  D_{\mu\nu} (p-k)
  \gamma^\mu \ S_F(k) \ \gamma^\nu,
\label{eq:SDeq}
\end{equation}
where  $i S_F^{-1}(p)=Z^{-1} (-p^2) (\xbar{p} - \Sigma(-p^2))$ and $i S^{-1}(p)=(\xbar{p} - m_0)$ are the full and bare fermion inverse propagators, respectively, and
$i D_{\mu\nu}(p)= (g_{\mu\nu} -p_\mu p_\nu/p^2)/p^2$ the bare gauge boson  propagator in the Landau gauge, and $C_2$ is  the quadratic Casimir of the fermion  of the gauge theory, with $C_2= (N_C^2 -1)/(2N_C)$ for 
  the fundamental  representation in $SU(N_C)$.   
After the angular integration,  the 
ladder SD equation in Landau gauge for $\Sigma (x\equiv -p^2)$ 
reads:
 \begin{equation}
 \Sigma(x) = m_0 +  
 \frac{3 C_2}{4\pi} \alpha \int^{\Lambda^2}
 dy\,   \left[\frac{\theta(x-y)}{x} + \frac{\theta(y-x)}{y}\right] \frac{y \Sigma(y)}{ y +\Sigma^2(y)},   \quad  (Z^{-1}(x) \equiv 1).  
 \label{SDeq}
 \end{equation}
 This form is reduced back to the form of the NJL gap equation with $\Sigma(x)\equiv m_F$,  Eq.(\ref{NJLgap}), if the kernel is local: $\theta(x-y)/x + \theta(y-x)/y \rightarrow  1/\Lambda^2$, such as in the case of the massive gauge boson,  $i D_{\mu\nu} \sim g_{\mu\nu}/\Lambda^2$ (See also Eq.(\ref{gaugedNJLSD})).  
 
Eq.(\ref{SDeq})  is converted into a differential equation plus IR and UV boundary conditions \cite{Fukuda:1976zb}:
\begin{eqnarray}
\left( x \Sigma(x) \right)'' 
+ \alpha \frac{3 C_2}{4 \pi} \frac{\Sigma(x)}{x + \Sigma^2(x)} &=& 0, 
\label{eq:diffSDo}\\
\lim_{x\rightarrow 0} x^2 \Sigma'(x) &=& 0,
\label{eq:IRBC}\\
\left.
 \left( x \Sigma(x) \right)'\right|_{x=\Lambda^2} &=& m_0.
\label{eq:UVBC}
\end{eqnarray}
The asymptotic solution of Eq.(\ref{eq:diffSDo}) at $x \gg \Sigma^2(x)$  takes the form 
$\Sigma (x)\sim  m_F (x/m_F^2)^a$, with a conventional normalization $\Sigma(x=m_F^2) =m_F$,  which is plugged back into the equation to yield 
$(a+1) a +\alpha (3C_2)/(4\pi) =0$, i.e., $a= (-1 \pm \sqrt{1-3C_2\alpha/\pi})/2$. 

For $\alpha<\frac{\pi}{3C_2}\equiv \alpha_{\rm cr}$, either solution, dominant ($a=(-1+\omega)/2$) or non-dominant ($a=(-1-\omega)/2$), has a power behavior, 
which does not satisfy the UV boundary condition Eq.(\ref{eq:UVBC}) for the chiral limit $m_0=0$, where $\omega \equiv \sqrt{1-\alpha/\alpha_{\rm cr}}$.
The solution exists only at the presence of the explicit breaking $m_0$, namely the explicit breaking solution with the renormalized mass 
$m_F=m_R= Z_m^{-1}\, m_0$, which yields the anomalous dimension $\gamma_m$ in the unbroken phase \cite{Leung:1985sn}:
\begin{equation}
m_0= m_R\left(\frac{\Lambda}{m_R}\right)^{-1+\omega},\quad \gamma_m =\Lambda  \frac{\partial \ln Z_m^{-1}}{\partial \Lambda} =1-\sqrt{1-\frac{\alpha}{\alpha_{\rm cr}}}<1\quad \left(\alpha<\alpha_{\rm cr}=\frac{\pi}{3 C_2}\right)\,.
\label{gammamconformal}
\end{equation}
The result is written in terms of the one-loop anomalous dimension $\gamma_m^{(one-loop)}=3C_2\alpha/(2\pi)$ as
$\gamma_m=1-\sqrt{1-2\gamma_m^{(one-loop)}}$ which coincides with $\gamma_m^{(one-loop)} $ for  $\alpha/\alpha_{\rm cr} \ll 1$: 

On the other hand, the 
SSB solution  does exist for
\begin{equation}
\alpha >\frac{\pi}{3C_2}=\alpha_{\rm cr}\,,
\end{equation}
where $a=(-1\pm i \tilde \omega)/2$ with $\tilde \omega\equiv \sqrt{\alpha/\alpha_{\rm cr}-1}$, and the solution is of the oscillating form
$\Sigma(x) \sim \frac{m_F^2}{\sqrt{x}}  \frac{1}{\tilde \omega} \sin \left(\tilde \omega [ \ln (\sqrt{x}/m_F)  +\delta] \right)$, $\delta ={\cal O} (1)$, which satisfies the UV boundary condition
as 
\begin{equation}
0=m_0\sim \frac{m_F^2}{\Lambda \tilde \omega } \sin \left(\tilde \omega \ln\left(\frac{4\Lambda}{m_F}\right)\right)\, ,
\end{equation}
for $\tilde \omega \ln\left(\frac{4\Lambda}{m_F}\right)=n \pi$ (numerically $e^\delta \simeq 4$).
In the large $N_C$ limit  the critical coupling itself is numerically small, $\alpha_{\rm cr}\sim 1/N_C \ll 1$, although the effective coupling in the
condensate channel is $C_2 \alpha_{\rm cr} ={\cal O} (1)$ as was the case in the NJL coupling. This is the reason why the ladder approximation 
yields reasonable result.

The ground state solution is $n=1$, which yields the dynamical mass $m_F$ of
the Berezinsky-Koterlitz-Thouless (BKT) form of essential-singularity (``Miransky scaling'') \cite{Miransky:1984ef}:
\begin{eqnarray}
m_F&\simeq& 4 \Lambda \exp\left(-\frac{\pi}{\sqrt{\frac{\alpha}{\alpha_{\rm cr}}-1}}\right)\,,\quad  \left(\alpha> \alpha_{\rm cr}=\frac{\pi}{3 C_2} \ne 0\right)\,,
\label{Miransky}\\
&=&0 \,,\quad \left(\alpha<\alpha_{\rm cr}\right)\,.
\end{eqnarray}
This is compared with the NJL gap equation Eq. (\ref{SSBNJL}) and D-dimensional NJL Eq.(\ref{SSBDNJL}), and also with the
BCS Eq.(\ref{BCSgap}) and 2-dimensional model Eq.(\ref{D2NJL}). Again the large hierarchy 
\begin{equation}
m_F\ll \Lambda\quad \left(\alpha/\alpha_{\rm cr} -1\ll 1\right)
\label{hierarchy}
\end{equation}
 can be realized near criticality (near chiral symmetry restoration). 
 
 The essential-singularity scaling yields a peculiar phase transition, dubbed ``conformal phase transition'' \cite{Miransky:1996pd}, which is different from the typical 2nd order phase transition
as  the Ginzburg-Landau phase transition. While the order parameter such as $m_F$ 
 is continuously changed as $m_F\ne0$ to $m_F=0$ from $\alpha>\alpha_{\rm cr}$ to $\alpha<\alpha_{\rm cr}$, the spectrum changes discontinuously,
 since there is no light spectrum in $\alpha<\alpha_{\rm cr}$ (conformal, unparticle) in contrast to the SSB phase where mass spectrum all
 goes to zero as $\alpha\rightarrow \alpha_{\rm cr}+0$. It reflects the fact that the essential singularity is not analytic at $\alpha=\alpha_{\rm cr}$.   
 The light spectrum is possible for $\alpha<\alpha_{\rm cr}$ only when $m_0\ne0$ which violates the conformality. Thus all the mass spectrum $M$ 
 for the conformal phase $\alpha<\alpha_{\rm cr}$ scales like the explicit breaking renormalized 
 mass $m_R$,
 which is given by Eq.(\ref{gammamconformal}) as  $M \sim m_R \sim m_0^{1/(1+\gamma_m)}$ \cite{Miransky:1998dh}, in conformity with the hyperscaling relation frequently used 
 in the lattice analyses for the conformal signals.

Eq.(\ref{Miransky})  implies that the coupling $\alpha$ is a function of $m_F/\Lambda$ with the nonperturbative beta function:
\begin{eqnarray}
\beta^{(NP)}(\alpha) &=&\Lambda \frac{\partial \alpha(\Lambda)}{\partial \Lambda}
= - \frac{2\pi^2\alpha_{\rm cr}}{\ln^3 (\frac{4\Lambda}{m_F})} 
=  - \frac{2\alpha_{\rm cr}}{\pi} \left(\frac{\alpha}{\alpha_{\rm cr}}-1\right)^{\frac{3}{2}} 
\label{NPbeta},\\
\alpha(\mu) &=&\alpha_{\rm cr} 
\left[1+ \frac{\pi^2}{\ln^2(\frac{4 \mu}{m_F})}
\right] \,,\label{NPrun}
\end{eqnarray}
with $\alpha_{\rm cr}$ being now regarded as a nontrivial ultraviolet fixed point (approached much faster than the asymptotic freedom $\sim 1/\ln \mu$). 
The asymptotic form of the SSB solution is $\Sigma (x) \sim m_F^2/\sqrt{x}$ which is compared with the Operator Product Expansion
$\Sigma(x) \sim \frac{m_F^3}{x} (\frac{x}{m_F^2})^{\gamma_m/2}$ to yield a large anomalous dimension: \cite{Yamawaki:1985zg,Bando:1986bg}
\begin{equation}
\gamma_m = 1\quad \left(\alpha>\alpha_{\rm cr}\right).
\end{equation}     
This ladder result is the characteristic feature of the walking technicolor.

Due to this mass generation which breaks the scale symmetry spontaneously, the ladder scale symmetry is also broken
explicitly producing the new nonperturbative trace anomaly besides the perturbative trace anomaly induced by the cutoff regularization $\Lambda$ (See Ref. \cite{Matsuzaki:2015sya} and references cited therein):
\begin{eqnarray}
 \langle\partial_\mu D^\mu \rangle= \langle\theta_\mu^\mu\rangle^{(NP)} 
  &\equiv&  \langle\theta_\mu^\mu\rangle^{(full)} -  \langle\theta_\mu^\mu\rangle^{(perturbative)}  =
 \frac{\beta^{(NP)}(\alpha)}{4\alpha}  \langle G_{\mu\nu}^2\rangle^{(NP)} 
\,,\nonumber \\
&\simeq& 
-  N_F N_C  \frac{4 \xi^2}{\pi^4} 
m_F^4 \,, \quad \left(\xi \simeq 1.1\right)\,,
\label{NPanomaly}
\end{eqnarray}
where  $\langle G_{\mu\nu}^2\rangle^{(NP)}
 \equiv \langle G_{\mu\nu}^2\rangle^{(full)}
 -\langle G_{\mu\nu}^2\rangle^{(perturbative)} $ is the nonpertubative gluon condensate and $N_F$ is a number of flavors of massless fermions
(besides color $N_C$). 
Note that although $ \frac{\beta^{(NP)}(\alpha(\mu))}{4\alpha(\mu)} $ and 
 $\langle G_{\mu\nu}^2\rangle^{(NP)}_{(\mu)} $ are depending on the renormalization point $\mu$, the trace anomaly 
$\langle\theta_\mu^\mu\rangle^{(NP)}$ is not as it should be (the energy-momentum tensor $\theta_{\mu\nu}$
 is a conserved current and is not renormalized), with both dependence being cancelled each other precisely. \cite{Matsuzaki:2015sya}
 
This ladder dynamics was the basis for the walking technicolor \cite{Yamawaki:1985zg,Bando:1986bg} where the coupling is almost non-running $\alpha(\mu) \approx \alpha_{\rm cr}$ for $m_F<\mu<\Lambda$
even after the SSB takes place to produce the nonperturbative running.

The  non-zero critical coupling also exists 
in the asymptotically free gauge theory including the QCD in a more sophisticated way, in spite of no explicit non-zero critical coupling separating the SSB phase and the non-SSB phase, namely the
 QCD is in one phase always in the SSB similarly to the BCS.  The theory is classically scale-invariant but actually has an intrinsic mass scale 
 $\Lambda_{\rm QCD}$ due to the trace anomaly by the quantum effects (regulator).  The intrinsic scale $\Lambda_{\rm QCD}$ is usually given by the one-loop beta function: $\Lambda_{\rm QCD} =\mu e^{-1/(b_0\alpha(\mu))}=\Lambda e^{-1/(b_0\alpha(\Lambda))}$ with $b_0$ given in Eq.(\ref{2loop}). 
 Although this looks like the BCS mass generation in Eq.(\ref{BCSgap}), 
 $\Lambda_{\rm QCD}$ should not be confused with the mass generation $m_F$.
 The existence of the intrinsic scale $\Lambda_{\rm QCD}$ does not necessarily
imply the SSB $m_F\ne 0$ as in the  NJL model where the intrinsic scale $1/\sqrt{G}$ does not necessarily imply the mass $m_F$. 

 The QCD coupling $\alpha(\mu)$ runs  depending on the renormalization scale $\mu$ in units of $\Lambda_{\rm QCD}$ to grow in the infrared region. 
  The  fermion mass $m_F$ is dynamically
 generated due to the fermion-antifermion condensate which takes place in the infrared region $\mu < m_F$ where the coupling becomes strong as to exceed
 the ``hidden'' critical coupling of order 1: $N_C \alpha(\mu<m_F) >N_C \alpha_{\rm cr} ={\cal O}(1)$. In the usual QCD it so happens that  $m_F ={\cal O} (\Lambda_{\rm QCD})$.
 In a wider parameter space, however,  we can see the cases $m_F=0$ (chiral restoration) and $m_F\ll \Lambda_{\rm QCD}$ (near chiral restoration), where the non-zero critical coupling is actually essential.
  
 The ``hidden'' non-zero critical coupling become ``visible'' when the system  is put in the medium with finite temperature $T$ and density with baryon chemical potential $\mu_B$, where the running coupling
 in the infrared region is no longer growing indefinitely and levels off  at  the relevant energy scale of order of $T$ or $\mu_B$. Then for $T,\mu_B$ such that $\alpha(T),\alpha(\mu_B) <\alpha_{\rm cr}$ the SSB would not take place, namely the chiral symmetry restoration occurs as has been studied actively. In contrast to the disappearance of the
 fermion-antifermion condensate, the BCS dynamics for fermion-fermion condensate instead can be operative in 
 the finite density   even with
 the weakest coupling due to the Fermi surface, which is called color superconductor.
 
 Here I discuss another case to visualize  the non-zero critical coupling in 
 the QCD-like vector-like $SU(N_C)$ gauge theory with  $N_F (\gg N_C)$ massless technifermions,  still in  the asymptotically free theory $N_F<11N_C/2$
with the running coupling vanishing  in the ultraviolet region. This is the basis for the walking technicolor to be discussed later and I denote the intrinsic scale $\Lambda_{\rm QCD}$ as $\Lambda_{\rm TC}$ hereafter.

\begin{figure}[h]
\begin{center}
\includegraphics[width=5.5cm]{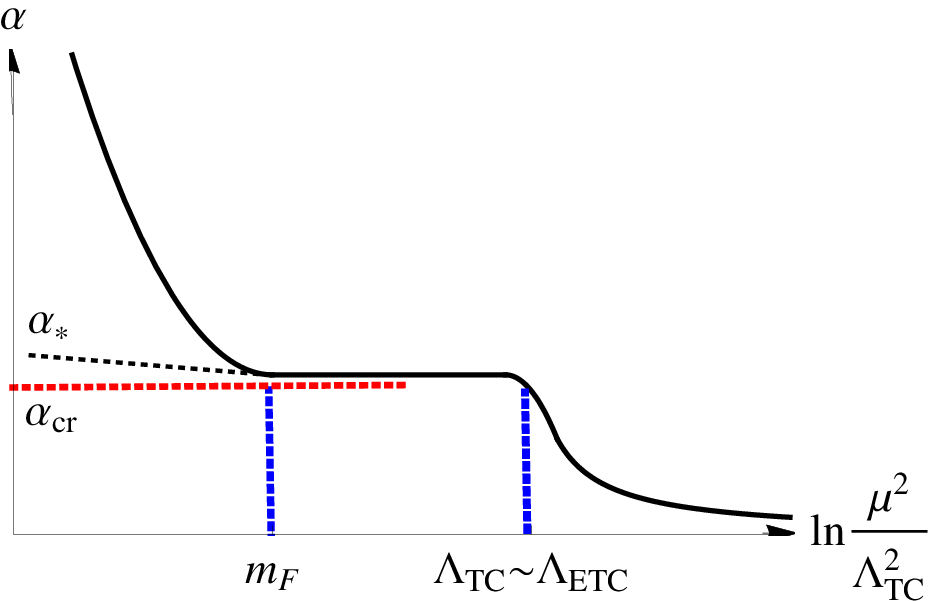} 
\hspace{15pt}
   \includegraphics[width=5.5cm]{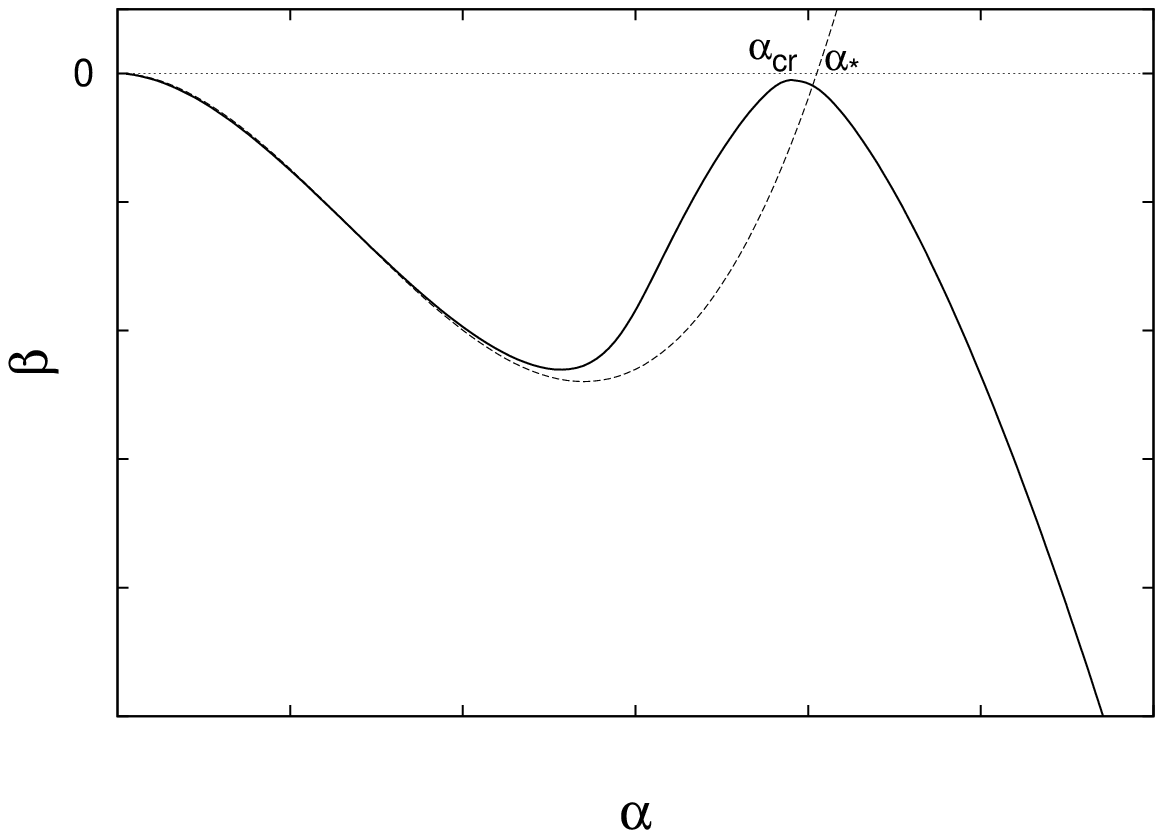} 
\vspace{15pt}
\caption{ Schematic picture of  the running coupling (left) and the beta function (right) in the region,
with the (quasi-)infrared fixed point $\alpha_*$ and the critical coupling as a (quasi-)ultraviolet fixed point $\alpha_{\rm cr}$.
Perturbative coupling in $\alpha<\alpha_{\rm cr}$ is smoothly connected to the nonperturbative one in the
region  $\alpha>\alpha_{\rm cr}$.
}
\label{beta:whole}
\end{center} 
 \end{figure}

 When one increases $N_F$, the vacuum polarization due to the
virtual fermion-antifermion pairs (loop effects) increases the screening of the charges in the long distance 
(infrared energy region), which is
 operative opposite to the asymptotically free anti-screening effects of the gluon loops: in the ultraviolet region $\mu\gg \Lambda_{\rm TC}$  the coupling is small and running is essentially one-loop dominated, while in the infrared region $\mu\ll \Lambda_{\rm TC}$ where  the coupling grows, the higher loop effects particularly by the fermion loop screening effects are getting  dominant,  which 
 then balances the anti-screening effects  to tend to make the coupling  level off. Then the dynamical mass $m_F$ such that  $\alpha(\mu=m_F) \simeq \alpha_{\rm cr}$ will be getting 
 smaller, as we increase $N_F/N_C$:
 \begin{equation}
 \frac{m_F}{\Lambda_{\rm TC}} \searrow \quad {\rm for} \quad \frac{N_F}{N_C} \nearrow\,,
 \end{equation}
  in contrast to the
 ordinary QCD with $m_F={\cal O}(\Lambda_{\rm QCD})$ for $N_F=N_C=3$. It then
  eventually could realize at certain large $r\equiv N_F/N_C\gg 1 $  an infrared fixed point $\alpha(\mu) < \alpha_*=\alpha(0)<\alpha_{\rm cr}$), which implies 
  that  no SSB takes place and no bound states exist (``unparticle''), the phase called ``conformal window''.\footnote{  Here we are talking about the phase transition in the parameter $N_F/N_C$ by changing the theory.
  It does not imply the existence of two phases in one theory with fixed $N_F/N_C$.  See the discussions below and 
  Fig. \ref{beta:whole}
    }
  The approximate scale symmetry is operative with almost nonrunning coupling in the infrared region $\mu\ll \Lambda_{\rm TC}$, although it is violated explicitly by $\Lambda_{\rm TC}$ due to 
  the trace anomaly 
  in the ultraviolet region $\mu \gg \Lambda_{\rm TC}$ where the coupling is 
  running as in  the usual asymptotically free theory.

    The existence of the conformal window in fact can been seen explicitly at two-loop beta function, which is scheme-independent while higher loops are not:
\cite{Caswell:1974gg}.  
\begin{eqnarray}
\beta^{(2-loop)}(\alpha) &=& -b_0 \alpha^2 -b_1 \alpha^3,\nonumber\\
b_0&=&\frac{1}{6\pi} (11N_C- 2N_F), \quad b_1= \frac{1}{24\pi^2}\left(34N_C^2-10N_C N_F -3 \frac{N_C^2-1}{N_C} N_F\right),\nonumber\\
 \alpha_*&=&\alpha_*(N_F,N_C) = -\frac{b_0}{b_1} \,,
 \label{2loop}
\end{eqnarray}
where we have $\beta^{(2-loop)}(\alpha=\alpha_*)=0$ by balancing the one-loop $-b_0 \alpha^2$ ($<0$ as far as asymptotically free, i.e., $N_F<11N_C/2$)  with the two-loop contributions $-b_1 \alpha^3>0$ at the infrared limit $\mu=0$, which is realized only when $b_1<0$ s.t.
$N_F\gg N_C$ is satisfied. Note that $\alpha_*\searrow $ as $N_F/N_C\nearrow$ and $\alpha_*=\alpha_*(N_F,N_C) $ exists for $N_F^*< N_F < 11N_C/2  $ ($N_F^*\simeq 8$ for $N_C=3$). 

In the context of large $N_C$ limit, such a situation corresponds to the 
``anti-Veneziano limit'' (in distinction to the original Veneziano limit with $N_F/N_C\ll 1$):\cite{Matsuzaki:2015sya}
\begin{equation} 
N_C \rightarrow \infty \quad {\rm and} \quad  N_C \cdot \alpha = {\rm fixed},
 \quad  {\rm with} \quad r\equiv N_F/N_C ={\rm fixed} \,\, \gg 1\,.
 \label{antiVeneziano}
\end{equation}
The anti-Veneziano limit in fact realizes a situation very close to the ladder approximation, with the $r=N_F/N_C$
behaving as a continuous parameter. 
Then the theory has two phases in the parameter space $r$:  SSB phase for $r>r_{\rm cr}$ such that $\alpha(\mu) >\alpha_{\rm cr}$ and the non-SSB phase otherwise.

 In the case $\alpha_*<\alpha_{\rm cr}$, there in fact exists no
SSB $m_F\equiv 0$ and no bound states (``unparticle''). The coupling is almost constant for all the infrared region $\mu<\Lambda_{\rm TC}$ (infrared conformality), while it is running in the ultraviolet region $\mu>\Lambda_{\rm TC}$ essentially
as the one-loop running, in accord with the scale symmetry violation due to the
perturbative trace anomaly. \footnote{For the region $\alpha(\mu)>\alpha_{\rm cr} >\alpha_*$, there might exist
SSB, in which case there might exist two phases separated by the ultraviolet fixed point at $\alpha_{\rm cr}$ in the sense similar to the conjecture on the asymptotically non-free gauge theory 
 such as the strong coupling QED for $\alpha(\mu) >\alpha_{\rm cr}$,
 although such a fixed point may be a Gaussian fixed point (trivial theory) \cite{Kogut:2005pm}.
}

On the other hand, for  $\alpha_* > \alpha_{\rm cr}$  the SSB takes place with mass $m_F$ generated similarly to the ladder SD result in Eq. (\ref{Miransky}) with $\alpha$ replaced by $\alpha_*$ \cite{Appelquist:1996dq}, 
\begin{equation}
m_F\sim \Lambda_{\rm TC} \exp \left( - \frac{\pi}{\sqrt{\frac{\alpha_*}{\alpha_{\rm cr}}-1}}\right)\quad \left( \ll \Lambda_{\rm TC}  \quad {\rm for} \quad \frac{N_F}{N_C}\quad  {\rm s.t.} \quad \frac{\alpha_*(N_F,N_C)}{\alpha_{\rm cr} (N_C)}-1 \ll 1 \right)\,, 
\end{equation}
where the phase transition in the parameter space $r$ has a characteristic essential singularity scaling (Miransky-BKT scaling), which
 takes the same type of the ``conformal phase transition'' 
as the ladder one \cite{Miransky:1996pd}.

Once $m_F$ is generated,  the scale symmetry is explicitly broken so as to yield the nonperturbative trace anomaly, Eq.(\ref{NPanomaly}), responsible for the nonperturbative running of the coupling, Eq.(\ref{NPrun}), and 
the would-be infrared fixed point $\alpha_*$ at two-loop is actually washed out. 
The resultant coupling would 
have a form with (quasi) ultraviolet fixed point $\alpha_{\rm cr}$ similarly to Eq.(\ref{Miransky}) for
$\alpha(\mu) >\alpha_{\rm cr}$ ($\mu <\Lambda_{\rm TC}$), while it still has a remnant of infrared fixed point 
(quasi fixed point) for $\alpha(\mu) <\alpha_{\rm cr}\simeq \alpha_*$ ($\mu>\Lambda_{\rm TC}$). 
Thus the theory is in one phase, which is not separated by $\alpha_{\rm cr}$. The beta function has no exact zero at
$\alpha_{\rm cr}$ and the coupling runs through $\alpha_{\rm cr}$ continuously. See Fig.\ref{beta:whole}.

 To summarize  the origin of mass in the  strong coupling theories,
 the mass $m_F$ originates  from the intrinsic scale $\Lambda$ 
 (Lagrangian parameter $G$ or the trace anomaly $\Lambda_{\rm QCD} (\Lambda_{\rm TC}))$ through SSB which takes place only in the strong coupling phase with the coupling larger than the non-zero critical coupling.
 There is no mass generation $m_F\equiv 0$ in the weak coupling phase, even though the theory has intrinsic scale $\Lambda$. 

\section{Hidden Symmetries  in the SM Higgs Lagrangian 
\cite{Fukano:2015zua}}

Here we recapitulate  Ref. \cite{Fukano:2015zua} to show that   the SM Higgs 
Lagrangian Eq.(\ref{Higgs}) in the form of the {\it linear sigma model}, Eqs.(\ref{sigma}) and (\ref{sigmaMatrix}), is rewritten into precisely the form equivalent to the {\it scale-invariant} version of the 
chiral $SU(2)_L\times SU(2)_R$ {\it nonlinear sigma model} based on the manifold $G/H$, with $G=SU(2)_L\times SU(2)_R$ and $H=SU(2)_{L+R}=SU(2)_V$,   as far as it is in the broken phase, with {\it both the  chiral and scale symmetries spontaneously broken} due to the same Higgs VEV
$v\ne 0$, and thus are {\it both nonlinearly realized}.  

The SM Higgs Lagrangian is further shown to be {\it gauge equivalent} to the {\it scale-invariant} 
version \cite{Kurachi:2014qma} of the Hidden Local Symmetry (HLS) Lagrangian \cite{Bando:1984ej,Bando:1987br,
Harada:2003jx}, which contains {\it possible new vector bosons}, analogues of the $\rho$ mesons,  as the gauge bosons of the (spontaneously broken) HLS {\it hidden behind the SM Higgs Lagrangian}.

Let us discuss the Higgs Lagrangian in the form of  Eqs.(\ref{sigma}) and (\ref{sigmaMatrix}):  
The potential minimum exists at the chiral-invariant circle:
\begin{equation}
\langle \sigma^2(x)\rangle = \frac{-\mu_0^2}{\lambda}\equiv v^2\,, \quad \sigma^2(x) \equiv  {\hat \sigma}^2(x) +{\hat \pi_a}^2(x)\,.
\label{chicircle}
\end{equation}
In Eq.(\ref{sigmaMatrix}) any complex matrix $M$ can be decomposed into the Hermitian (always diagnonalizable) matrix $H$  and unitary matrix $U$ as $M=HU$ ( ``polar decomposition'' ): 
 \begin{equation}
 M(x) = H(x)\cdot U(x)\,, \quad H(x)=\frac{1}{\sqrt{2}} \left(\begin{array}{cc}
 \sigma(x) & 0\\
 0  &\sigma(x)
 \end{array}\right)
 \,, \quad U(x)= \exp\left(\frac{2i \pi(x)}{F_\pi}\right)  \,,
 \label{Polar}
 \end{equation}
 with $\pi(x)=\pi^a(x) \frac{\tau^a}{2} \,(a=1,2,3)$ and $F_\pi=v=\langle  \sigma(x) \rangle$. The chiral transformation of $M$ is inherited by $U$,\footnote{The nonlinear realization was first introduced by K. Nishijima (then at Osaka City University) \cite{Nishijima} (in $G/H =U(1)_L\times U(1)_R/U(1)_V$ case) to make the nucleon massive in a chiral invariant way using the NG boson $\pi$ as 
 $M_N  {\bar \Psi}_L  \Psi_R$, where the physical (massive) nucleon $\Psi_{L/R} =(\xi^\dagger \psi_L, \xi \psi_R)$
  transforms as $\Psi_{L/R} \rightarrow h(\pi(x),g_{L/R}) \cdot \Psi_{L/R}$, ($h\in H$), while the original nucleon  field
 $\psi_{L/R}$ does as $\psi_{L/R} \rightarrow g_{L/R} \cdot \psi_{L/R}$. Here the nonlinear base $(\xi^\dagger, \xi)$ is defined by $U(x)=\xi^2(x)$ with the transformation, $(\xi^\dagger,\xi)=(e^{-i \pi/F_\pi},e^{i\pi/F_\pi}) \rightarrow h(\pi(x),g_{L/R}) \cdot (\xi^\dagger,\xi) \cdot g_{L/R}^\dagger$, $h^\dagger \, h=1$, in accord with Eq.(\ref{transformation}). 
 See, e.g., Ref.\cite{Bando:1987br}.
 } while $H$ is a chiral singlet such that:
 \begin{equation}
 U \rightarrow g_L \, U\, g_R^\dagger\,,\quad H \rightarrow H\,,
 \label{transformation}
 \end{equation}
where $g_{L/R} \in SU(2)_{L/R}$ and $U \, U^\dagger=1$ implies $\langle U\rangle =\langle  \exp\left(\frac{2i \pi(x)}{F_\pi}\right)\rangle=1 \ne 0$, namely the spontaneous breaking of the chiral symmetry is taken granted in the polar decomposition.
 Note that {\it the radial mode $\sigma$ is a chiral-singlet in contrast to $\hat \sigma$ which is a chiral non-singlet} transformed 
 into the chiral partner ${\hat \pi}_a$ by the chiral rotation. The {\it physical particles are $\sigma$ and $\pi$} which are
defined by the  nonlinear realization, {\it in contrast to the tachyons $\hat \sigma$ and ${\hat \pi}_a$}.

 We further parametrize $\sigma(x)$  as 
 \begin{equation} 
 \sigma(x) =v \cdot \chi(x)\,,\quad \chi(x)=\exp\left(\frac{\phi(x)}{F_\phi}
 \right)
 \,,
 \label{NLscale}
 \end{equation}
 where $F_\phi=v$ is the decay constant of the dilaton $\phi$ as the Higgs. 
 The scale (dilatation) transformations for these fields are 
 \begin{equation}
 \delta_D \sigma =(1 +x^\mu \partial_\mu) \sigma \,, \qquad  
\delta_D \chi=(1+x^\mu \partial_\mu) \chi\,, \qquad 
\delta_D \phi= F_\phi+x^\mu \partial_\mu\phi\,. 
 \end{equation}
Note that $\langle  \sigma(x)\rangle= v \langle \chi(x) \rangle = v\ne 0$ breaks spontaneously the scale symmetry, but not the chiral symmetry, since 
$ \sigma(x)$ ($\chi(x)$ as well) is a chiral singlet.
This is a nonlinear realization of the scale symmetry: 
the $\phi(x)$ is a dilaton, NG boson of the spontaneously broken scale symmetry. Although $\chi$ is a dimensionless field,
it transforms as that of dimension 1, while $\phi$ having dimension 1 transforms as the dimension 0, instead.
 
 Plugging Eqs.(\ref{Polar})  and (\ref{NLscale}) into the SM Higgs Lagrangian Eq.(\ref{sigmaMatrix}),     
 we straightforwardly arrive at the SM Higgs Lagrangian in the striking form:\cite{Fukano:2015zua}
 \begin{eqnarray}
  {\cal L}_{\rm Higgs}
 &=&\left[ \frac{F_\phi^2}{2} \left(\partial_\mu \chi \right)^2+ \frac{F_\pi^2}{4}{\chi}^2\cdot {\rm tr} \left(\partial_\mu U \partial^\mu U^\dagger\right)\right]
  - V(\phi)\nonumber\\
 &=&\chi^2(x) \cdot \left[ \frac{1}{2} \left(\partial_\mu \phi\right)^2  +\frac{F_\pi^2}{4}{\rm tr} \left(\partial_\mu U \partial^\mu U^\dagger\right)\right] -V(\phi)\nonumber \,,\\
 V(\phi)&=& \frac{\lambda}{4} v^4 \left[\left(\chi^2(x) -1\right)^2-1\right]  =\frac{M_\phi^2 F_\phi^2}{8} \left[\left(\chi^2(x) -1\right)^2-1\right] \,,
 \label{SNLSM}
 \end{eqnarray}
which  is nothing but {\it the scale-invariant nonlinear sigma model} with $F_\phi=F_\pi=v$, an effective theory of the walking technicolor~\cite{Matsuzaki:2012mk,Matsuzaki:2015sya},  apart from the form of the explicit scale-symmetry breaking potential $V(\phi)$ (see Eq.(\ref{WTC})).

The explicit scale-symmetry breaking comes only from the  potential $V(\phi)$ such that $\delta_D V(\phi) = \lambda v^4\chi^2=-\theta_\mu^\mu$  whose scale dimension $d_{\theta}=2$ (originally the tachyon mass term)  instead of 4 of the walking technicolor: namely, the scale symmetry is broken only by the dimension 2 operator.\footnote{Note that {\it mass term of all the SM particles except the Higgs is scale-invariant}. By the electro-weak gauging as usual; $\partial_\mu U\Rightarrow {\cal D}_\mu U= \partial_\mu U -i g_2 W_\mu U +i g_1 U B_\mu$
 in Eq.(\ref{SNLSM}), we see that the mass term of $W/Z$ is scale-invariant thanks to the dilaton factor $\chi$, and so is the mass term of the SM fermions $f$: $g_Y  \bar f h f
 =(g_Y v/\sqrt{2}) (\chi \bar f f)$, all with the scale dimension 4.
 }
This yields the mass of the (pseudo-)dilaton as the Higgs $M_\phi^2=2\lambda v^2$, which is in accord  with the Partially Conserved Dilatation Current (PCDC) for $\partial^\mu D_\mu=\theta_\mu^\mu$:
\begin{equation}
M_\phi^2 F_\phi^2=-\langle0|\partial^\mu D_\mu|\phi\rangle F_\phi=-d_{\theta} \langle \theta_\mu^\mu\rangle =2\lambda v^4\langle \chi^2(x) \rangle=2\lambda v^4\,,
\label{PCDC}
\end{equation}
with $F_\phi=v$, where $D_\mu$ is the dilatation current: $\langle 0|D_\mu(x) |\phi\rangle=-i q_\mu F_\phi e^{-i q x}$, or equivalently $\langle 0| \theta_{\mu\nu}|\phi(q)\rangle= F_\phi (q_\mu q_\nu- q^2 g_{\mu\nu}/3)$.  
 
Hence the SM Higgs
as it stands is a (pseudo) dilaton, with the {\it mass arising from the dimension 2 operator}
 in the potential, which vanishes for $\lambda\rightarrow 0$:
 \begin{equation}
 M_\phi^2=2\lambda v^2 
 \rightarrow 0
 \quad \left(\lambda\rightarrow 0\,, \,\,
 v=\sqrt{\frac{-\mu^2_0}{\lambda}} = 
 {\rm fixed}\,\ne 0\right)
 \label{conformallimit}
  \end{equation}
  (``conformal limit''\cite{Fukano:2015zua}).\footnote{With vanishing  potential, $V(\phi) \rightarrow 0$, this limit still gives an  {\it interacting theory where the physical particles $\pi$ and $\phi$ have derivative coupling} in the same sense as in the nonlinear chiral Lagrangian Eq.(\ref{NLS}). It should be contrasted to the triviality limit, $\lambda\rightarrow 0$ {\it without fixing $ v=\sqrt{\frac{-\mu^2_0}{\lambda}}\ne 0$}, which yields only a  free theory of  tachyons $\hat \pi$ and $\hat \sigma$. This limit should also be distinguished from the popular limit $\mu^2_0\rightarrow 0 $ with $\lambda=$fixed $\ne 0$, where the Coleman-Weinberg potential as the explicit scale symmetry breaking is generated by the trace anomaly  (dimension 4 operator) due to the quantum loop. }
In fact the Higgs mass 125 GeV implies that the SM Higgs is in near conformal limit with $v=$ fixed: 
\begin{equation}
\lambda=\frac{1}{2} \left(\frac{M_\phi}{v}\right)^2 \simeq \frac{1}{2} \left(\frac{125\,{\rm GeV}}{246\,{\rm GeV}}\right)^2 \simeq \frac{1}{8} \ll 1\,.
\end{equation}  
It should be noted that {\it $\lambda\ll 1$ (with $v =$ fixed $\ne 0$) can be realized even when the underlying theory is strong coupling}, particularly when the {\it scale symmetry is operative},
 as we discuss later in both NJL type theory, Eq.(\ref{NJLconformallimit}), and the strong coupling gauge theory (walking technicolor) in the anti-Veneziano limit,  Eq.(\ref{weakcouplingTD}).

On the other hand, if we take the limit $\lambda \rightarrow \infty$, 
then the SM Higgs Lagrangian goes over to the  usual nonlinear sigma model {\it without scale symmetry}:  
\begin{equation}
{\cal L}_{{\rm NL}\sigma}=\frac{F_\pi^2}{4} {\rm tr} \left(\partial^\mu U\partial_\mu U^\dagger\right)\,,\quad \left(\lambda\rightarrow \infty\,, \,\,
 v=\sqrt{\frac{-\mu^2_0}{\lambda}} = 
 {\rm fixed}\,\ne 0\right)
 \label{NLS}
\end{equation}
where the potential is decoupled with $\chi(x)$ frozen to the minimal point $\chi(x)\equiv 1$ ($\phi(x) \equiv  \langle \phi(x) \rangle=v\ne 0$), so that the scale symmetry breaking is transferred from the potential  to the kinetic term, 
which is no longer transformed as the dimension 4 operator.  This is known to be a good effective theory (chiral perturbation theory) of the ordinary QCD which in fact lacks the scale symmetry at all,
perfectly consistent  with
the nonlinear sigma model, Eq.(\ref{NLS}). However, it cannot be true for  
the walking technicolor which does have the scale symmetry, and the effective theory must respect the symmetry of the underlying theory, in a form of the scale-invariant nonlinear sigma model Eq.(\ref{SNLSM}) in the conformal limit $\lambda\rightarrow 0$.   

Once rewritten in the form of Eq.(\ref{SNLSM}),  it is easy to see \cite{Fukano:2015zua}
that  the {\it SM Higgs Lagrangian is gauge equivalent to the ``scale-invariant HLS model'' (s-HLS)}\cite{Kurachi:2014qma}, a scale-invariant version of the HLS model \cite{Bando:1984ej,Bando:1987br,Harada:2003jx} \footnote{
The s-HLS model was also discussed in a different context, ordinary QCD in medium.\cite{Lee:2015qsa}
}
, which {\it contains massive spin-1 states}, spontaneously broken HLS gauge bosons, as  possible yet other composite states  in some
underlying theory hidden behind the 
SM Higgs.

The HLS can be made explicit by dividing $U(x)$ into two parts:
\begin{equation}
 U(x)= \xi_L^\dagger(x) \cdot \xi_R(x)\,,
 \label{U:decomp}
\end{equation}  
where $\xi_{R,L}(x)$ transform under $G_{\rm global} \times H_{\rm local}$ as
\begin{equation}
\xi_{R,L}(x) \rightarrow h(x) \cdot \xi_{R,L}(x) \cdot { g^\prime}_{R,L}^\dagger\,,\quad 
U(x) \rightarrow {\hat g}_L U(x) { g^\prime}_R^\dagger \quad \quad
 \left(h(x)\in H_{\rm local},\, {g^\prime}_{R,L}\in G_{\rm global} \right)
\end{equation}
The $H_{\rm local} $ is a gauge symmetry of group $H$ arising from the redundancy (gauge symmetry) how to divide
$U$ into two parts. Then we can introduce the HLS gauge boson $V_\mu(x)$ by covariant derivative as 
\begin{equation}
D_\mu \xi_{R,L}(x) = \partial_\mu \xi_{R,L} (x)-i V_\mu(x) \xi_{R,L}(x) \,,
\label{HLScovariant}
\end{equation}
which transform in the same way as $\xi_{R,L}$. Then we have covariant objects transforming homogeneously under $H_{\rm local}$: 
  \begin{eqnarray}
 {\hat \alpha}_{\mu,R,L}(x)&\equiv& \frac{1}{i}D_\mu \xi_{R,L}(x) \cdot \xi_{R,L}^\dagger(x) =\frac{1}{i}\partial_\mu \xi_{R,L}(x) \cdot \xi_{R,L}^\dagger(x) 
 - V_\mu(x)\,, \nonumber\\ 
 {\hat \alpha}_{\mu, ||,\perp}(x)&\equiv & \frac{1}{2}\left({\hat \alpha}_{\mu,R}(x) \pm  {\hat \alpha}_{\mu, L}(x)\right) 
\nonumber \\  
&=& 
\Bigg\{ \begin{array}{c}
  \frac{1}{2i} \left(\partial_\mu \xi_{R}(x) \cdot \xi_{R}^\dagger(x)+ \partial_\mu \xi_{L}(x) \cdot \xi_{L}^\dagger(x) \right)- V_\mu(x)= {\alpha}_{\mu,||}(x) - V_\mu(x)\\  
 \frac{1}{2i} \left(\partial_\mu \xi_{R}(x) \cdot \xi_{R}^\dagger(x)- \partial_\mu \xi_{L}(x) \cdot \xi_{L}^\dagger(x)  \right) = {\alpha}_{\mu,\perp}(x)  \end{array}\nonumber\,,\\
 \{  {\hat \alpha}_{R,L}(x), {\hat \alpha}_{||,\perp}(x)\}
  &\rightarrow& h(x) \cdot \{ {\hat \alpha}_{R,L}(x), {\hat \alpha}_{||,\perp}(x)\}\cdot h^\dagger(x)\,.  
      \end{eqnarray}
We thus have two independent invariants under the larger symmetry $G_{\rm global} \times H_{\rm local}$:
\begin{equation}
\,
{\cal L}_A=v^2\cdot {\rm tr} {\hat \alpha}_{\perp}^2(x)\,,\quad 
 \,\quad
{\cal L}_V=v^2\cdot  {\rm tr} \,{\hat \alpha}_{||}^2(x) = v^2\cdot   {\rm tr} \, \left(V_\mu(x) - {\alpha}_{\mu,||}(x)\right)^2 \,. 
\end{equation} 
Hence the scale-invariant version of the Higgs Lagrangian, Eq.(\ref{SNLSM}), in the conformal limit $\lambda \rightarrow 0\,, v={\rm fixed}$, can be extended to the scale-invariant version having the HLS (s-HLS):\cite{Kurachi:2014qma}
\begin{equation}
{\cal L}_{\rm s-HLS} = \chi^2(x) \cdot \left(\frac{1}{2} \left(\partial_\mu \phi\right)^2 + {\cal L}_A+ a {\cal L}_V\right)\,,
\label{SHLS}
\end{equation} 
with $a$ being an arbitrary parameter. 

We now fix the gauge of HLS as $\xi_L^\dagger=\xi_R=\xi=e^{i \pi/v}$ such that $U=\xi^2$. 
Then  $H_{\rm local}$ and $H_{\rm global} (\subset G_{\rm global})$ 
get simultaneously broken spontaneously (Higgs mechanism), leaving the 
diagonal subgroup $H=H_{\rm local}+H_{\rm global}$, which is nothing but the subgroup $H$ of the original $G$ of $G/H$: $H\subset G$. 
According to the Higgs mechanism, the HLS gauge boson $V_\mu(x)$ acquires the mass $\frac{1}{2} a (g_H \, v)^2\,(V^a_\mu(x))^2$ through the invariant ${\cal L}_V$ after rescaling the kinetic term of $V_\mu$ by the HLS gauge coupling $g_H$ as 
$V_\mu(x) \rightarrow g_H\, V_\mu(x)$. {\it Obviously the vector boson mass terms are scale-invariant thanks to the nonlinear realization of the scale symmetry!} 
For the low energy $p^2<M_V^2$ where the kinetic term can be ignored, the HLS gauge boson $V_\mu$ becomes just an auxiliary field to be solved away to yield ${\cal L}_V=0$. 
 Noting that  ${\cal L}_A=v^2 \cdot{\rm tr} 
{\hat \alpha}_{\mu,\perp}^2(x)=v^2  \cdot{\rm tr}{\alpha}_{\mu,\perp}^2(x)=  \frac{v^2}{4}\cdot {\rm tr} \left(\partial_\mu U \partial^\mu U^\dagger\right) $ by a straightforward algebraic calculation, we see that 
${\cal L}_{\rm s-HLS}$ in Eq. (\ref{SHLS})  is simply reduced back to the original SM Higgs Lagrangian ${\cal L}_{\rm Higgs}$ in nonlinear realization, Eq.(\ref{SNLSM}). Note that {\it the HLS gauge boson acquires the
scale-invariant mass term thanks to the dilaton factor $\chi^2$}, the nonlinear realization of the scale symmetry, in sharp contrast to the {\it Higgs (dilaton)  which acquires mass only from the explicit breaking of the scale symmetry}.

The electroweak gauge bosons ($\in {\cal R}_\mu ({\cal L}_\mu)$) are introduced by extending the covariant derivative of Eq.(\ref{HLScovariant}) 
this time by gauging $G_{\rm global}$, which is {\it independent of $H_{\rm local}$} in  the HLS extension:
\begin{equation}
D_\mu \xi_{R,L}(x)\Rightarrow {\hat D}_\mu \xi_{R,L}(x)\equiv  \partial_\mu \xi_{R,L} (x)-i V_\mu(x) \, \xi_{R,L}(x)  +i \xi_{R,L}(x)\, {\cal R}_\mu ( {\cal L}_\mu)\,.
\label{fullcovariant}
\end{equation}
As usual in the Higgs mechanism, the gauge bosons of ${\rm gauged-}H_{\rm global} (\subset 
{\rm gauged-}G_{\rm global}$) get mixed with the gauge bosons of HLS, leaving  only the gauge bosons of the unbroken diagonal subgroup $({\rm gauged-}H)=H_{\rm local} + ({\rm gauged-}H_{\rm global})$ be massless after mass diagonalization.
 We then finally have a gauged s-HLS version of the Higgs Lagrangian (gauged-s-HLS):
 \begin{equation}
 {\cal L}^{\rm gauged}_{\rm s-HLS}=  \chi^2(x)\cdot 
 \left[
\frac{1}{2} \left(\partial_\mu \phi\right)^2  + {\hat {\cal L}}_A+ a {\hat {\cal L}}_V
  \right]\,, 
  \label{gaugedsHLS}
  \end{equation}
 with 
 \begin{equation}
  {\hat {\cal L}}_{A,V}= {\cal L}_{A,V} \left( D_\mu \xi_{R,L}(x) \Rightarrow {\hat D}_\mu \xi_{R,L}(x)\right) .
     \end{equation}

The new HLS boson may be identified with the LHC diboson events  \cite{Aad:2015owa}, with the model parameter choice
consistent with the reported results \cite{MY16}, 
similarly to the walking technirho which is also the HLS
boson for the large chiral symmetry ($SU(8)_L\times SU(8)_R$ in the one-family model) \cite{Fukano:2015hga,Fukano:2015zua}.

A salient feature of the new vector boson of HLS in the scale-invariant SM Lagrangian is that 
the scale-invariant  vector boson mass terms in Eq. (\ref{gaugedsHLS}) having the $\phi$ (Higgs $H$) field in the overall conformal factor $\chi^2(x)$ yield 
 $\phi$ couplings only to the diagonal pairs of the (longitudinal) SM gauge bosons $H-W/Z-W/Z$ or of those of the new vector bosons $H-V-V$
  after the mass diagonalization (as it should be done).    
Thus the new HLS vector bosons hidden in the SM only couple to the Higgs in a pair of themselves as $V-V-H$ but not 
in  the off-diagonal combination with the SM weak bosons $W/Z$: 
\begin{equation} 
  V- W/Z - H \, {\rm coupling} \, = \, 0   \,,
  \label{Conformalbarrier}
\end{equation} 
namely, the decay $V\rightarrow W/Z + H$ is forbidden by
 the scale/conformal symmetry ({\it Conformal Barrier})~\cite{Fukano:2015zua}, in sharp contrast to the
 popular ``equivalence theorem'' which implies comparable coupling of $V$ to $W/Z + W/Z$ and $W/Z +H$, based on  the usual 
 (non scale-invariant) Higgs field identification $\hat \sigma=v+ H$, with the Higgs $H$ being
 on the same footing as the NG modes $\hat \pi$ which are the longitudinal modes of $W/Z$ by the equivalence theorem. 
Consequently, the $V$ predominantly decays to the weak boson pairs $WW/WZ$. 
In other words, the popular consequence of the ``equivalence theorem''  is invalidated by the scale/conformal symmetry.
The absence of $V\to WH/ZH$ signatures at the LHC Run-II thus could indirectly probe the existence of the (approximate) scale/conformal invariance of the system involving
$V$, $W, Z$ and $H$.

It is straightforward to extend the internal symmetry group to $G_{\rm global}$ =$SU(N_F)_L\times SU(N_F)_R$ and $H_{\rm local}= SU(N_F)_V$. 
The Lagrangian then takes the form 
\begin{equation} 
 {\cal L}_{\rm s-HLS} 
 = \chi^2(x) \cdot  \left( 
 \frac{1}{2} \left(\partial_\mu \phi\right)^2
 + 
F_\pi^2 \left[ 
{\rm tr}[ \hat{\alpha}_{\mu \perp}^2 ] 
+ 
a \, {\rm tr}[ \hat{\alpha}_{\mu ||}^2 ] 
\right] 
\right) 
+ \cdots  
\,, \label{sHLS}
\end{equation}  
where $F_\pi$ is related to $v=246$ GeV as $F_\pi = v/\sqrt{N_F/2}$. 
This form of the Lagrangian  is the same as that of the effective theory of the one-family ($N_F=8$) walking technicolor \cite{Kurachi:2014qma}, except for the shape of the scale-violating 
potential $V(\phi)$  which has a scale dimension 4 (trace anomaly) in the case of the walking technicolor instead of  2 of the SM Higgs case (Lagrangian mass term). We shall come back to this later.
 
\section{Strong Coupling NJL as  the UV completion of the Weak Coupling SM Higgs Lagrangian \cite{Yamawaki:2015tmu}}
\label{NJLvsSM}

  Let us now recapitulate Ref. \cite{Yamawaki:2015tmu} which elaborated  the composite Higgs model based on the strong coupling theory $G>G_{\rm cr}\ne 0$ pioneered by Nambu.
In the NJL  model \cite{Nambu:1961tp} for the $N_C-$component 2-flavored fermion $\psi$ the Lagrangian takes the form:
\begin{eqnarray}
{\cal L}_{\rm NJL} &=& \bar \psi i\gamma^\mu\partial_\mu \psi + \frac{G}{2} \left[ (\bar \psi \psi)^2 +(\bar \psi i \gamma_5 \tau^a \psi)^2\right]\nonumber\\
&=& \bar \psi \left(i\gamma^\mu\partial_\mu  +\hat \sigma +i\gamma_5 \tau^a \hat \pi_a \right)\psi -\frac{1}{2G} \left({\hat \sigma}^2 +{\hat \pi_a}^2\right)\,,
\label{NJL}
\end{eqnarray}
where the equations of motion of the auxiliary fields  $\hat \sigma \sim G \bar \psi \psi$ and ${\hat \pi}^a \sim G \bar \psi i\gamma_5 \tau^a\psi$ are plugged back into the Lagrangian to get the original Lagrangian.
In the large $N_C$ limit ($N_C \rightarrow \infty$ with $N_C G\ne 0$ fixed), after rescaling the induced kinetic term to the canonical one, $Z_\phi^{1/2}  \hat \sigma \rightarrow \hat \sigma$,
 the quantum theory for $\hat \sigma$ and $\hat \pi$ sector  yields 
precisely the same form as the SM Higgs Eq.(\ref{Higgs}), 
with: \cite{Eguchi:1976iz}
\begin{eqnarray}
\mu_0^2&=& \left(\frac{1}{G} - \frac{1}{G_{\rm cr}}\right) Z_\phi^{-1} =-2m_F^2=- v^2 Z_\phi^{-1} =- \lambda v^2 < 0 \quad (G>G_{\rm cr}=\frac{4\pi^2}{N_C \Lambda^2} )
\nonumber\\
\lambda&=& Z_\phi Z_\phi^{-2} = Z_\phi^{-1} = \left[\frac{N_C}{8\pi^2} \ln \frac{\Lambda^2}{m_F^2}\right]^{-1} \sim \left[\frac{N_C}{8\pi^2} \ln \frac{\Lambda^2}{v^2}\right]^{-1} \,,
\label{tachyon}
\end{eqnarray}
where the gap equation has been used:
\begin{equation}
\frac{1}{G} - \frac{1}{G_{\rm cr}}= -\frac{N_C}{4\pi^2} m_F^2 \ln \frac{\Lambda^2}{m_F^2}=- 2 m_F^2 Z_\phi =-F_\pi^2=-v^2\,.
\label{gap}
\end{equation}

Eq.(\ref{tachyon}) shows that the {\it tachyon with $\mu_0^2<0$ is in fact generated} by the dynamical effects  for the {\it strong coupling}  $G>G_{\rm cr}\ne 0$,
 corresponding to the generation of mass $m_F\ne0$ in the gap equation.
Or, we can explicitly see it by computing the $\bar \psi \psi$ bound state  using the  $m_F=0$  solution  (wrong solution) of the gap equation at $G>G_{\rm cr}$. The correct spectrum $M_{\pi}^2=0, M_{\phi}^2 =2 \lambda v^2 =-2\mu_0^2=4m_F^2$ can be 
obtained when we use the correct solution $m_F\ne 0$ in the gap equation. 
The last equality $M_\phi^2=4 m_F^2$, often dubbed ``BCS mass relation'',  is specific to the $N_C \rightarrow \infty$ (with $N_C\, G\ne 0$ fixed) limit of the NJL model ({\it ``weak coupling'' limit}  $G >G_{\rm cr} \sim 1/N_C\rightarrow 0$ in the {\it strong coupling phase}),  but not the general outcome of the NJL model nor the generic  linear sigma model.

There are two extreme limits for $\lambda$ in Eq.(\ref{tachyon}) : 
$\lambda \rightarrow0$ ($N_C\gg 1$ and/or $\Lambda/v^2 \gg 1$) reproduces precisely the conformal limit, or scale-invariant nonlinear sigma model limit, Eq.(\ref{SNLSM}), of the SM Higgs Lagrangian, 
while $\lambda\rightarrow \infty$ ($N_C, \Lambda^2/v^2 ={\cal O}(1)$) does  the nonlinear sigma model limit without scale symmetry, Eq.(\ref{NLS}).

We are particularly interested in the limit 
\begin{equation}
\lambda =
\frac{1}{\frac{N_C}{8\pi^2} \ln \frac{\Lambda^2}{v^2}}
\rightarrow 0
\label{NJLconformallimit}
\end{equation}
 (conformal limit in Eq.(\ref{conformallimit})), which is 
realized in the strong coupling theory with $G>G_{\rm cr}\ne 0$ for $\Lambda/v\rightarrow \infty$ and/or $N_C\rightarrow \infty$, with $v=F_\pi=F_\phi \ne 0$ fixed.\footnote{
If $\Lambda$ is regarded as a physical cutoff in contrast to the nonperturbative renormalization arguments below, this argument would not be realistic for the 125 GeV Higgs with $\lambda\simeq 1/8$, corresponding to $\Lambda\simeq v\cdot e^{32\pi^2/N_C} \gg 10^{19}$ GeV. For the NJL model with $N_D$ doublets, however, we would have $\Lambda\simeq v \cdot e^{32\pi^2/(N_D N_C)} \sim 10^{11}$ GeV for $N_D=N_C=4$, somewhat realistic if the condensate is mainly triggered by the strong 
(ETC-induced) four-fermion coupling rather than the technicolor gauge coupling \cite{Miransky:1988gk,Matumoto:1989hf}
 in the one-family technicolor model \cite{Dimopoulos:1979sp} ($N_F=2 N_D=8$, see later discussions.).
}
 Then the effective Lagrangian in the large $N_C$ limit takes precisely the same as the SM Higgs Lagrangian, 
which is further equivalent to the scale-invariant nonlinear sigma model, Eq.(\ref{SNLSM}), as mentioned before. Now the SM
Higgs is identified with the composite (pseudo-)dilaton  
with mass vanishing  $M_\phi^2 =2 \lambda v^2 \rightarrow 0$.

The limit theory gives an {\it interacting (nontrivial) low energy effective theory even in the $\Lambda/v \rightarrow \infty$ limit}: a scale-invariant nonlinear sigma model ~\cite{Matsuzaki:2012mk,Matsuzaki:2015sya} where massless $\pi$ and $\phi$ are {\it interacting with each other} with 
the (derivative) couplings $\sim (1/F_\pi, 1/F_\phi) \ne 0$.  It is actually the basis for the {\it scale-invariant chiral perturbation theory} (sChPT) with the derivative expansion as a loop expansion \cite{Matsuzaki:2013eva}, although the Yukawa couplings of
$\pi,\phi$ to the fermions are vanishing $g_Y\sim m_F/F_\pi, m_F/F_\phi \rightarrow 0$ (The composite particles are still interacting due to the loop divergence compensation of the vanishing Yukawa coupling). 
This limit should be sharply distinguished from a similar limit $\Lambda/m_F \rightarrow \infty$, $m_F=$fixed (not $\Lambda/v \rightarrow \infty$, $v=$fixed), which is  the famous  triviality limit (Gaussian fixed point) where the theory becomes a free theory: free massive scalar for $G<G_{\rm cr}$ and free tachyon for $G>G_{\rm cr}$, with not just the Yukawa couplings but all the couplings vanishing. 

One might wonder why dilaton in NJL model? Obviously the NJL model has  the explicit scale-breaking coupling $G$ having dimension $[M]^{-2}$. But this scale is an ultraviolet scale to which the low energy effective theory
is insensitive. This is in exactly the same sense as in the scale-invariant ladder gauge theory, Eqs.(\ref{Miransky}) and (\ref{hierarchy}),  where the intrinsic scale $\Lambda_{\rm TC}$ generated by the trace anomaly can be far bigger than the infrared scale of spontaneous
symmetry breaking $F_\pi, F_\phi ={\cal O}  (v) \ll \Lambda_{\rm TC}$ thanks to the approximate scale symmetry due to the almost nonrunning coupling.  

In fact we can formulate the nonperturbative running of the (dimensionless) four-fermion coupling $g=\frac{\Lambda^2}{4\pi^2} G$ in the same way as the Miransky nonperturbative renormalization: The gap equation Eq. (\ref{gap}) reads
\begin{equation}
\left(\frac{1}{g_{\rm cr}}  - \frac{1}{g}\right) \Lambda^2 =N_C m_F^2 \ln \frac{\Lambda^2}{m_F^2} \simeq 4\pi^2 v^2\,,\quad g_{\rm cr}=\frac{1}{N_C}\,,
\label{gap2}
\end{equation}
which leads to  a nonperturbative beta function for $g>g_{\rm cr}$:
\begin{equation}
\beta (g) = \Lambda \frac{\partial g(\Lambda)}{\partial \Lambda}\Bigg|_{v=\rm fixed}=-\frac{2}{g_{\rm cr}} g\cdot  (g-g_{\rm cr})\,,\quad g(\mu) = g_{\rm cr}\frac{1}{
1-\frac{4\pi^2 g_{\rm cr} v^2}{\mu^2} 
}
\label{NJLbeta}
\end{equation}
by {\it fixing $v=$ constant} (instead of the conventional limit with $m_F=$ constant) 
and taking $\Lambda \rightarrow \infty$. 
Thus, without troublesome log factor,  $g=g_{\rm cr}=1/N_C$ is the ultraviolet fixed point defining a nontrivial interacting theory in the continuum limit.  As the running coupling $g(\mu)$ 
reaches $g_{\rm cr}$ even much faster than the walking coupling in Eq.(\ref{NPrun}),
the scale symmetry is operative 
$g(\mu) \approx g_{\rm cr}$ for 
the wide region $4\pi^2 g_{\rm cr}v^2\ll \mu^2< \Lambda^2$.

 We also have  $- \langle \bar \psi_i \psi_j \rangle = \delta_{i,j} \Lambda^2 m_F N_C/(4\pi^2)= Z_m^{-1}\,\delta_{i,j} v^3 N_C /(4\pi^2) $, where $Z_m^{-1}=Z_m^{-1}(\Lambda/v)= (\Lambda/v)^2 [N_C \ln (\Lambda^2/v^2)/(4\pi^2)]^{-1/2} $ is the mass renormalization constant, which
 implies\footnote{
 Hence the operators have  scale dimension $d_{\bar \psi \psi}= 1$ and $d_{(\bar \psi \psi)^2}= 2$ in the large $N_C$ limit.
Note that $\gamma_m$ is actually slightly smaller by the $1/\ln(\Lambda^2/v^2)$ than  ``$\gamma_m=2$''  in the conventional limit $m_F=$ constant, so that $d_{(\bar \psi \psi)^2}$ is slightly larger than 2 actually, i.e., possible eight-fermion operators corresponding to the $\lambda \phi^4$ would have dimension $d_{(\bar \psi \psi)^4} >4$, barely
irrelevant. This is contrasted to the conventional limit where  the eight-fermion
operators  are marginal and hence the NJL coupling without such counter terms would not be renormalizable nor interacting theory in the continuum limit (See the section 8 of the 3rd entry of Ref.\cite{Kondo:1991yk}). 
 }  
\begin{equation}
\gamma_m = Z_m \Lambda \frac{\partial Z_m^{-1}}{\partial \Lambda} =2 - 1/\ln(\Lambda^2/v^2) 
\longrightarrow 2 \quad (\Lambda/v \rightarrow \infty)\,.
\label{gamma}
\end{equation}
Thus we may write $\bar \psi_i \psi_j= -Z_m^{-1}\delta_{i,j} v^3 N_C/(4\pi^2) \cdot \chi $, or $(G/2)(\bar \psi \psi)^2=2 g N_C\Lambda^2 v^2\cdot \chi^2/[\ln(\Lambda^2/v^2)]$. 
The gap equation implies $\beta(g)/g = -g (4\pi^2) (v^2/\Lambda^2)$.
Putting all together, we have $\beta(g)/g\cdot (G/2) \cdot
(\bar \psi \psi)^2|_{g\rightarrow g_{\rm cr}=1}= - \lambda v^4\chi^2$. 
Then we get the {\it explicit scale symmetry breaking in the dimension 2 operator}: $\theta_\mu^\mu= \frac{\beta(g)}{g} \frac{G}{2} \left[
\left(\bar \psi \psi\right)^2 +\left(\bar \psi i\gamma_5 \tau^a \psi \right)^2
\right]= -\lambda v^4 \chi^2$, 
where $\lambda = 8\pi^2/[N_C\ln(\Lambda^2/v^2)]\rightarrow 0$ as in Eq.(\ref{tachyon}).

The PCDC follows in precisely the same way as in the SM Higgs as $M_\phi^2 F_\phi^2 =-d_{(\bar \psi \psi)^2} \langle \theta_\mu^\mu\rangle =2 \lambda v^4$ (See below Eq.(\ref{SNLSM})).
In any case the trace of energy-momentum tensor vanishes in the limit $\lambda \sim 1/[N_C\ln (\Lambda^2/v^2)] \rightarrow 0$, and  the dilaton mass should come from the trace anomaly in the $1/N_C$ sub-leading loop effects, or the chiral loops of the effective theory Eq.(\ref{SNLSM}).  

Again the spin 1 composites can also be introduced via HLS, precisely in the same way as Eq.(\ref{SHLS})  for the SM Higgs 
Lagrangian. This time it can be done more explicitly by introducing the vector/axialvector type four-fermion coupling which
in fact become the ``explicit'' composite HLS gauge bosons.(See section 5.3 of Ref.\cite{Bando:1987br}).

Incidentally, the above prescription to have an interacting nontrivial continuum theory of the NJL model 
is similar to the renormalizability arguments of the $D$-dimensional NJL model ($2<D<4$) \cite{Kikukawa:1989fw} 
  and the gauged NJL model \cite{Kondo:1991yk} both without troublesome log factor, 
  although in the case at hand the explicit scale-breaking from the Lagrangian parameters, i.e., the  four-fermion interaction and fermion mass term (if present), depend on the renormalization point (vanish at the UV limit). 

The beta function Eq.(\ref{NJLbeta}) and anomalous dimension Eq.(\ref{gamma})  of the  $D-$dimensional four-fermion theory renormalizable in $1/N_C$ expansion are given: \cite{Kikukawa:1989fw}\begin{equation}
\beta(g) =- \frac{D-2}{g_{\rm cr}} \, g\,(g-g_{\rm cr})\,,\quad \gamma_m=(D-2)\frac{g}{g_{\rm cr}} \rightarrow D-2 \,\,(g\rightarrow g_{\rm cr})\,,
\label{DNJLbetagamma}
 \end{equation} 
which follows from the gap equation Eq.(\ref{SSBDNJL}) and the condensate $\langle \bar \psi \psi\rangle \sim (\Lambda/m_F)^{D-2}$, respectively, while the renormalization can be done independently of the phase  in this theory. 

For $D=2$  the ultraviolet fixed point $g=0$ and the infrared fixed point $g=g_{\rm cr}$
coincide, i.e.,  
\begin{equation}
\beta(\alpha)\rightarrow  -2 N_C g^2,\quad \gamma_m\rightarrow 2 N_C g,
\label{2NJLbetagamma}
\end{equation}
 which also follows directly from Eq.(\ref{D2NJL}). The essential singularity scaling in Eq.(\ref{D2NJL}) and the associated colliding ultraviolet and infrared fixed points   at $D=2$ are
 characteristic features of the conformal phase transition \cite{Miransky:1996pd} similarly to the BKT-Miransky scaling, where there exist no light bound states for $g< g_{\rm cr}=0$, while the bound states
 have mass of order of ${\cal O} (m_F) \rightarrow 0$ ($g\searrow g_{\rm cr}=0$) for $g>g_{\rm cr}$.   

It is also compared with the beta function of the  
gauged NJL model with walking gauge coupling $\alpha(\mu)=\alpha=$ constant: \cite{Kondo:1991yk}
\begin{equation}
 \beta(g)=\frac{\partial g}{\partial \ln \Lambda}\Bigg|_{\alpha, m_F}= - 2 N_C \left(g-g^{(+)}(\alpha)\right) \left(g-g^{(-)}(\alpha) \right)\,,
 \label{gaugedNJLbeta}
 \end{equation}
where \cite{Kondo:1988qd,Yamawaki:1988na} 
\begin{equation}
 N_C \, g^{(\pm)}(\alpha)=\frac{1}{4} \left(1\pm \omega\right)^2\,, \quad \omega \equiv \sqrt{1-\frac{\alpha}{\alpha_{\rm cr}}}\quad \left(0<\alpha<\alpha_{\rm cr}\right)\,.
 \label{criticalline}
 \end{equation}
 This follows from the gap equation (for $m_0=0$): 
 \begin{equation}
 \Sigma(x) = \frac{N_C G}{4\pi^2} \int^{\Lambda^2} d x \frac{x \Sigma(x)}{x+\Sigma(x)} +
 \frac{3 C_2}{4\pi} \alpha \int^{\Lambda^2}
 dy\,   \left[\frac{\theta(x-y)}{x} + \frac{\theta(y-x)}{y}\right] \frac{y \Sigma(y)}{ y +\Sigma^2(y)},  
 \label{gaugedNJLSD}
 \end{equation}
 which is a combined gap equation of Eq.(\ref{NJLgap}) and Eq.(\ref{SDeq}).
 Were it not for the NJL interaction $G=0$, the SSB solution would exist only for $\alpha>\alpha_{\rm cr}$ as we have explained in section 3. However, due to strong NJL coupling,
this time  it has the SSB solution for $g=G\Lambda^2/(4\pi^2) > g^{(+)}(\alpha)$ even at $\alpha<\alpha_{\rm cr}$: \cite{Kondo:1988qd,Yamawaki:1988na}
   \begin{equation}
m_F^{2\omega} = \Lambda^{2\omega} \left(
\frac{g-g^{(+)}(\alpha)}{g-g^{(-)}(\alpha)} \right) \,,\quad \left(g>g^{(+)}(\alpha)=g_{\rm cr}(\alpha)\,, \quad 0<\alpha<\alpha_{\rm cr}\right)\,
\label{gaugedNJLsol}
\end{equation} 
  as well as $\alpha>\alpha_{\rm cr}$, while $g<g^{(+)}(\alpha)$ is 
 the unbroken phase, $m_F= 0$,
 where the criticality is now extended to the critical line $g_{\rm cr}(\alpha)=g^{(+)}(\alpha)$  in the  $(g,\alpha)$ space instead of just the gauge coupling $\alpha$. 
 The beta function Eq.(\ref{gaugedNJLbeta})  is readily obtained from Eq.(\ref{gaugedNJLsol}), and
 the anomalous dimension is given as:~\cite{Miransky:1988gk, Kondo:1991yk}
  \begin{equation}
  \gamma_m =2N_C g +\frac{\alpha}{2\alpha_{\rm cr}}\,, \quad
 \gamma_m^{(\pm)} =  \gamma_m\Bigg|_{g=g^{(\pm)}(\alpha)}=1\pm \omega= 1 \pm \sqrt{1-\frac{\alpha}{\alpha_{\rm cr}}} \,.
 \label{gaugedNJLgamma}
  \end{equation} 
  
The critical line $g=g_{\rm cr}(\alpha)=g^{(+)}(\alpha)$, coming from the four-fermion coupling additional to the gauge dynamics (e.g., ETC coupling in the technicolor case), behaves as an ultraviolet fixed point, while the non-critical line $g^{(-)}(\alpha)$, coming from the four-fermion coupling induced by the gauge dynamics itself in a Wilsonian sense, does as  an infrared fixed point, both colliding ${\bar g}_{\rm cr}=g^{(+)}(\alpha)=g^{(-)}(\alpha)=1/(4 N_C)$ at $\alpha=\alpha_{\rm cr}$,
where the 
 SD gap  equation yields the Miransky-BKT scaling of the essential-singularity form (conformal phase transition):
\begin{equation}
m_F =\Lambda \exp \left( - \frac{g}{ g -{\bar g}_{\rm cr}} \right)\,, \quad \left(\alpha=\alpha_{\rm cr},  \, g>{\bar g}_{\rm cr}=\frac{1}{4 N_C}\right)\,.
\label{gNJLgapcritical}
\end{equation} 
Accordingly, the 
beta function and the anomalous dimension read 
\begin{equation}
\beta(g)= - 2 N_C \left(g-{\bar g}_{\rm cr}\right)^2\,, \quad \gamma_m=2 N_C g+\frac{1}{2} \quad \left(\gamma_m\Big|_{{\bar g}_{\rm cr}}=1\right),
\label{gNJLbetacritical}
\end{equation}
which is compared with Eq.(\ref{2NJLbetagamma}).

  The outstanding feature of the gauged NJL model with $0<\alpha \leq\alpha_{\rm cr},\, (1\leq\gamma_m<2)$ is the renormalizability
(in the sense of nontriviality, or no Landau pole) independently of the phase \cite{Kondo:1991yk}, when the gauge coupling is walking, $\alpha(\mu^2) \approx {\rm const.}$. (Similar results are obtained for the gauged Yukawa model \cite{Kondo:1993ty}.)  
The four-fermion operators have the full dimension $2< d_{(\bar \psi \psi)^2} = 2(3-\gamma_m) = 4-2\omega \leq4$ (relevant operator, or super renormalizable), including $d\simeq 2(1+ A/\ln \mu^2)>2 (d \ne 2)$
with a  moderately ``walking'' 
small coupling $\omega \simeq 1 -\frac{\alpha}{2\alpha_{\rm cr}} \simeq 1-\gamma_m \,\,(\gamma_m (\mu) \sim A/\ln \mu^2) $ with $A=18 C_2/(11 N_C -2 N_F) >1$ \footnote{For the NJL gauged by the actual QCD with 6 flavors ($u,\dots, t$), $A=8/7>1$, as in the original top quark condensate model \cite{Miransky:1988xi} satisfies
this renormalizability condition \cite{Kondo:1991yk}, though the electroweak gauge interaction invalidates it. 
If the SM gauge groups are embedded into a GUT which is usually walking $A>1$, then the GUT-gauged NJL is renormalizable, having the interacting continuum limit.
}
, in sharp contrast to the pure (non-gauged) NJL model with $\gamma_m=2,\, d_{(\bar \psi \psi)^2}=2$,  which is  a trivial theory having a Landau pole in the conventional way of the continuum limit keeping $m_F=$ constant  (not the way described in the above keeping $F_\pi=$ constant).\footnote{
For renormalizability of the gauge theories and gauged NJL model  in $D>4$ dimensions, with the extra dimensions $\delta=D-4$ compactified, see
Ref.\cite{Hashimoto:2000uk}.
}

\section{Top Quark Condensate a la NJL dynamics, the simplest UV completion of the SM Higgs}
\label{TopQuarkCondensate}

\begin{figure} [h]
 \begin{center}
  \includegraphics[width=6cm]{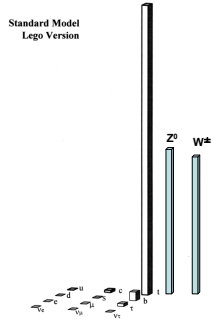}
\caption{Lego version of the Standard Model. Top quark mass vs. other masses in linear scale. }
\label{fig:SMLego}
\end{center} 
 \end{figure}

One of the concrete composite Higgs models as the 
straightforward
application of the NJL type theory is the top quark condensate model (Top-Mode Standard Model) \cite{Miransky:1988xi,Nambu:1989jt,Bardeen:1989ds}. 
The model predicted  that only the top quark among SM fermions has mass on the order of the weak scale (Fig.\ref{fig:SMLego}), at the time many people expected the top mass below 50 GeV. 

The explicit four-fermion Lagrangian of the top quark condensate takes the form: \cite{Miransky:1988xi}
\begin{eqnarray}
{\cal L}_{\rm MTY}&=&
G^{(1)} \left({\bar \psi}_L^i \psi_R^j\right)\left({\bar \psi}_R^j \psi_L^i\right) 
+\left[G^{(2)}\left({\bar \psi}_L^i \psi_R^j\right)\left(i\tau_2\right)^{ik} \left(i\tau_2\right)^{jl} 
\left({\bar \psi}_L^k \psi_R^l\right)+h.c.\right]\nonumber\\
&+&
G^{(3)}  \left({\bar \psi}_L^i\psi_R^j\right)
\left( \tau_3\right)^{jk} \left({\bar \psi}_R^k \psi_L^i\right)\,, \quad \left(G^{(i)}= g^{(i)} \frac{4\pi^2}{\Lambda^2}\right)\,.
\label{MTY}
\end{eqnarray}
The inclusion of other generations is straightforward \cite{Miransky:1988xi}. 
In the realistic case the SM gauge interaction, particularly QCD, was included via the gauged NJL model \cite{Miransky:1988xi}, 
where the critical coupling is actually the critical line  of the gauged NJL, $G_{\rm cr}(\alpha)=  g_{\rm cr}(\alpha)\cdot  (4\pi^2/\Lambda^2)$ with 
$ g_{\rm cr}(\alpha)$  in Eq.(\ref{criticalline}), 
while the  $U(1)_Y$ coupling is numerically negligible (the chiral gauge $SU(2)_L$ does not contribute to the condensate channel).  
The crucial ingredient of the model is again the {\it non-zero critical coupling}  {\it in sharp contrast to the ``bootstrap symmetry breaking'' \cite{Nambu:1989jt} 
based on the weakly-coupled BCS theory which has $G_{\rm cr}=0$} as already mentioned:
{\it only the top quark coupling is strong coupling larger than the critical coupling}
$G_t =G^{(1)}+G^{(3)} > G_{\rm cr}(\alpha)$ while $G_b=G^{(1)}-G^{(3)}$ and all others are less as well, $G_{c,s,d,u} <G_{\rm cr}(\alpha)$, so that only the top acquires the dynamical mass of order of weak scale  ${\cal O}(v)$ to produce
only three NG bosons to be absorbed into the $W/Z$ bosons \cite{Miransky:1988xi,Bardeen:1989ds}. 

We disregard $G^{(2)}$ and $G_b$  terms for the moment, then
Eq.(\ref{MTY}) simply reads:
\begin{eqnarray}
{\cal L}_{\rm MTY}&=& G_t \left( {\bar \psi}_L t_R\right) \left({\bar t}_R \psi_L\right) \nonumber \\
&=&   \bar \psi_L h t_R +h.c.  - \frac{1}{G_t}  h^\dagger h \,,
\end{eqnarray}
where $h=G_t \bar \psi_L t_R$ with $\hat \sigma \sim G_t \bar t  t, {\hat \pi}^0\sim G_t \bar t i\gamma_5 t, \hat \pi^{\pm}\sim G_t \bar t i\gamma_5 b, G_t
\bar b i\gamma_5 t$. This simplified version was also considered in \cite{Bardeen:1989ds}.

The effective theory of the pure bosonic sector at $1/N_C$ leading order is precisely the same as the SM Higgs Lagrangian Eq.(\ref{Higgs}) \cite{Bardeen:1989ds} 
as already mentioned in Section \ref{NJLvsSM},
  which happens to have the $SU(2)_L \times SU(2)_R$ global symmetry
not just the $SU(2)_L \times U(1)_R$ to be gauged by the electroweak gauge bosons. (The $SU(2)_R$ is explicitly broken only by the Yukawa term:     
$\bar  \psi_L  h t_R$.)
Then the effective theory of the top quark condensate model is nothing but the SM Higgs as the scale-invariant HLS model, Eq.(\ref{SHLS}), which includes the 3 NG bosons $\pi^{\pm,0}$ to be absorbed into $W,Z$ and
the Higgs $\phi(x)$ as a pseudo-dilaton, in the nonlinear realization in Eq.(\ref{Polar}) and (\ref{NLscale}), and in addition  the vector composites (``top-mode rho meson'') 
which can be identified \cite{MY16}  with the 2 TeV diboson events at LHC \cite{Aad:2015owa}.

If we further assume a small bottom condensate $\langle \bar b b \rangle$ by fine-tuning the bottom four-fermion coupling: $G_t>G_b >G_{\rm cr}(\alpha)$,
 then we have
a Peccei-Quinn type axion (``top-mode axion'') \cite{Miransky:1988xi,Luty:1990bg}, which acquires mass from the $G^{(2)}$ term in Eq.(\ref{MTY}).
(It may be identified with the 750 GeV diphoton
events at LHC \cite{750}.)

 The obvious phenomenological problem of the top-mode SM is the prediction of the top mass in the large $N_C$ limit relation (BCS mass relation)  to the Higgs mass:
 $M_\phi=2 m_t$.
which is actually modified by the effects of the SM gauge interactions,
$M_\phi \simeq \sqrt{2} m_t$ \cite{Bardeen:1989ds, Shuto:1989te}.  It is further modified to $M_\phi\simeq m_t$ 
by some of the non-leading order in $1/N_C$ expansion \cite{Bardeen:1989ds} using the ultraviolet boundary condition, ``compositeness condition'' \cite{Bardeen:1989ds}, at say the GUT
scale, 
where both effective couplings of top Yukawa $g_Y^t$ and the Higgs quartic coupling $\lambda$ diverge. 

Instead of the compositeness condition, we may consider the renormalizability of the gauged NJL model mentioned in the previous section.
A possible renormalizable top quark condensate model would then
be to unify the SM gauge interactions into a walking GUT, ``Top-mode walking GUT'' \cite{Yamawaki:1996tj}, 
which determines the values of the top Yukawa coupling $g_Y^t$
and the Higgs coupling $\lambda$  in terms of the GUT gauge coupling $g_{\rm GUT}$ all at the GUT scale $\Lambda_{\rm GUT}$
as the Pendleton-Ross infrared fixed point of the effective theory of the GUT-gauged NJL model, typically as ${g_Y^t}^2(\Lambda_{\rm GUT})\simeq \lambda(\Lambda_{\rm GUT}) \simeq \frac{3}{2} \,g_{\rm GUT}^2(\Lambda_{\rm GUT})$, instead of the diverging couplings of the compositeness condition. This generally yields prediction of  mass of $m_t$ and $M_\phi$ much smaller than that of  compositeness condition. 

Another possibility would be to include the near marginal operators: since the anomalous dimension
is very close  to 2, the  four-fermion operators (corresponding to the bare $\lambda \phi^4$ term) have the dimension $d\simeq 2$, so that formally irrelevant
eight-fermion operators could be near
marginal and compete with the higher order corrections in $1/N_C$ expansion, which may change the mass ratio
substantially. It is not known presently whether or not the relation $M_\phi \simeq m_t/2$ can be naturally realized by yet 
higher order effects in $1/N_C$ as well as the eight-fermion operators.

Yet other different solutions have been considered,  see e.g. 
top seesaw \cite{Dobrescu:1997nm}, and its NG boson Higgs version \cite{Fukano:2013aea} where the Higgs is a pseudo NG boson
living  in the coset space of the larger internal symmetry $G/H=U(3)\times U(1)/[U(2)\times U(1)^\prime]$ rather than the pseudo-dilaton.  

The LHC Run II will tell us whether or not the basic idea of the top quark condensate is on the right track.

\section{Walking Technicolor and Technidilaton}

Yet another composite Higgs model in the spirit of Nambu is the strong coupling gauge theory, similar in some sense to the 
QCD, a simple scale-up version dubbed technicolor \cite{TC}. However the original technicolor was excluded long time ago for
the problem of the Flavor-Changing Neutral Currents (FCNC) in a way to give mass to the SM fermions through four-fermion interaction from Extended Technicolor (ETC) \cite{Dimopoulos:1979es}
or some composite technicolor models \cite{Yamawaki:1982tg}. As the strong coupling gauge theory with {\it more explicit role of the non-zero critical coupling}, the walking technicolor \cite{Yamawaki:1985zg,Bando:1986bg} was proposed as a solution to the FCNC problem by a large anomalous dimension $\gamma_m =1$, based on the SSB solution of the scale-invariant dynamics, the 
ladder SD equation, and at the same time predicted a light composite Higgs, dubbed technidilaton,  as a pseudo NG boson of the approximate scale symmetry.\footnote{
Similar work on the FCNC problem \cite{Holdom:1984sk} was also done without notion of the anomalous dimension,  the scale symmetry,
nor the technidilaton.
Solving FCNC by a large anomalous dimension was proposed earlier \cite{Holdom:1981rm}, based on a pure assumption of the existence of  a gauge theory 
having nontrivial ultraviolet fixed point.
}

Here I recapitulate the explanation \cite{Matsuzaki:2015sya} on how the
technidilaton is a {\it naturally light and weakly coupled composite Higgs} out of {\it strongly coupled} underlying conformal gauge theory, the walking technicolor, in the light of the anti-Veneziano limit 
Eq.(\ref{antiVeneziano}). 
The technidilaton particularly for the one-family technicolor, with $N_F=8$ and $N_C=4$ \cite{Dimopoulos:1979es}, as the walking technicolor  is nicely fit to the current 125 GeV Higgs data at LHC \cite{Matsuzaki:2012mk,Matsuzaki:2012xx,Matsuzaki:2015sya}.

As we discussed in section \ref{SCGT}, ladder approximation is realized by the anti-Veneziano limit Eq.(\ref{antiVeneziano}) in  the large $N_F$ QCD. The evaluation of  the nonperturbative trace anomaly in the anti-Veneziano limit can be essentially given 
by the ladder result Eq.(\ref{NPanomaly}).
It then yields the mass $M_\phi$ and decay constant $F_\phi$ of the technidilaton $\phi$ 
through PCDC \cite{Bando:1986bg}  as in Eq.({\ref{PCDC}), this time in terms of the dimension 4 operator:\cite{Matsuzaki:2015sya,Hashimoto:2010nw}  
 \begin{eqnarray}
 M_\phi^2 F_\phi^2&=&  
-d_\theta \langle \theta_\mu^\mu \rangle 
=-  \frac{\beta^{(\rm NP)}(\alpha (\mu^2))}{\alpha(\mu^2)}
\, \langle G_{\mu \nu}^2(\mu^2)\rangle 
\simeq N_C N_F\frac{16 \xi^2}{\pi^4} m_F^4\quad \left(d_\theta=4 \right)
\label{PCDC2}\\
&\simeq& 2.5\left[\frac{8}{N_F}\frac{4}{N_C}\right] v^4 \,. \quad \left(v= 246\,{\rm GeV}\right)
\label{PCDC3}
\end{eqnarray}

First, the rightmost value in Eq.(\ref{PCDC2}) can be obtained by two different ladder calculations: 
one through direct evaluation of the vacuum energy by the effective potential at the stationary point (Solution of the SD equation, $\Sigma=\Sigma_{\rm sol}$)\cite{Gusynin:1987em}, $E=V_{\rm eff} (\Sigma =\Sigma_{\rm sol}) =\langle\theta^0_0\rangle
=(1/4)\langle\theta^\mu_\mu\rangle$,  the other  through the ladder evaluation  of the trace anomaly \cite{Hashimoto:2010nw,Matsuzaki:2015sya}, i.e., the  technigluon condensate $\langle G_{\mu\nu}^2\rangle$ times the nonperturbative beta function
Eq.(\ref{NPbeta}), both in precise agreement with each other. The agreement is  in highly nontrivial manner, being  {\it independent of the renormalization point $\mu$} as it should be:
 $\langle G_{\mu \nu}^2(\mu^2)\rangle \sim \ln^3 (\mu^2/m_F^2)$,  while $\beta^{({\rm NP})}  (\alpha (\mu^2)) /\alpha(\mu^2) \sim  1/\ln^3 (\mu^2/m_F^2)$, precisely cancelled by each other.\cite{Matsuzaki:2015sya}
 
Second, Eq.(\ref{PCDC3}) is obtained by use of the Pagels-Stokar formula:
\begin{equation}
 v^2=(246\,  {\rm GeV})^2 
= N_D F_\pi^2 \simeq 
N_F N_C\frac{\xi^2}{4\pi^2} \, m_F^2   
\simeq  m_F^2 \left[\frac{N_F}{8}\frac{N_C}{4}\right],
\label{PS}
\end{equation}
and the result indicates {\it important $N_F,N_C-$ dependence of $M_\phi^2F_\phi^2$ in the anti-Veneziano limit when $v=$ fixed} \cite{Matsuzaki:2015sya}.
Since the technidilaton is a flavor-singlet bound state, its decay constant by definition scales like $F_\phi^2 \propto N_F N_C m_F^2\, (\propto v^2)$ (Actually $F_\phi^2 \simeq N_F N_C m_F^2$). Then 
 $M_\phi^2/F_\phi^2, M_\phi^2/v^2 \sim 1/(N_F N_C) \rightarrow 0$ in the anti-Veneziano limit, where the technidilaton becomes NG boson although no exact
 massless limit exists: the situation is  in the same sense as the 
 $\eta^\prime$ meson in the original Veneziano limit $N_C \rightarrow \infty$ with $N_C\, \alpha=$fixed, and $N_F/N_C \ll 1$.\footnote{
 There exists no exact massless limit in the conformal phase transition at $\alpha=\alpha_{\rm cr}$ (this time $r(=N_F/N_C)= r_{\rm cr}$), with $m_F=0$, where no massless spectrum exists for $\alpha\leq \alpha_{\rm  cr}$ (conformal phase),
 in sharp contrast to the Ginzburg-Landau phase transition where the spectrum continuously passes through the phase transition point with massless particles. \cite{Miransky:1996pd} 
 }
 
 Although the PCDC relation together with  Pagels-Stokar formula does not give $M_\phi$ and $F_\pi$ separately,
 Eq.(\ref{PCDC3}) well accommodates numerically the desired result:
 \begin{equation}
 F_\phi\simeq 5 \,v \quad {\rm for} \,\, M_\phi \simeq \frac{v}{2} \simeq 125\,{\rm GeV}\quad \left(N_F=8, N_C=4\right)\,,
 \label{dilatonHiggs}
 \end{equation} 
in the one-family model, which is best fit to the current LHC data of the 125 GeV Higgs up to 30 \% uncertainty due to limitation of the ladder approximation \cite{Matsuzaki:2012mk,Matsuzaki:2015sya}.
Similar results are also obtained within 30\% uncertainty in the holographic model for the walking technicolor.\cite{Matsuzaki:2012xx}

Incidentally, at the criticality $\alpha=\alpha_{\rm cr}=\frac{\pi}{3 C_2}$, the anomalous dimension $\gamma_m=1$ implies that  the induced four-fermion interaction 
generated by the walking technicolor coupling itself (i.e., not the ETC-like gauge interaction additional to  the technicolor interaction) also becomes 
a marginal operator with $d=4$. Then the phase diagram should be considered in the wider coupling space
$(\alpha, g)$\cite{Leung:1985sn}, where $g$ is the dimensionless coupling of the induced four-fermion interaction in the form of
the gauged NJL model. See Eq.(\ref{gNJLgapcritical}) and (\ref{gNJLbetacritical}). Then we can predict  $M_\phi$ independently of 
$F_\phi$: the denominator of the renormalized $\sigma$ propagator $D^\sigma(p)$ can be evaluated at $p=0$ in the large $N_C$ limit  \cite{
Nonoyama:1989dq}:
 \begin{equation}
M_\phi^2=D^\sigma(0)=\frac{16\xi^2}{\pi^4}m_F^2\simeq \left(\frac{m_F}{2}\right)^2 \quad \left(\alpha=\alpha_{\rm cr}\,,\, g\searrow  {\bar g}_{\rm cr}\right)\,,
\end{equation}
which yields $M_\phi \simeq \frac{v}{2}\simeq 125$ GeV (!) through the Pagels-Stokar formula Eq.(\ref{PS}) for $N_F=8, N_C=4$, i.e., $v\simeq m_F$,  
and  in turn predicts 
$F_\phi^2 \simeq N_F N_C m_F^2$, 
combined with  the PCDC relation Eq.(\ref{PCDC2}). The result is quite consistent  with  Eq.(\ref{dilatonHiggs}) for the 125 GeV Higgs.

The effective theory of the walking technicolor with $N_F$ massless flavors takes {\it precisely the same scale-invariant form as the nonlinearly realized SM Higgs Lagrangian} in Eq.(\ref{SNLSM}), with $U=e^{i \pi^a\, T^a}$ being $N_F\times N_F$ unitary matrix (${\rm tr} \,T^a=0\,, {\rm tr} (T^aT^b)=\delta^{ab}/2,\,a=1,\cdots, N_F^2-1$), except that 
the explicit scale breaking comes from the different potential $V^{(4)}(\phi)$~\footnote{
This potential is indeed obtained by the explicit ladder computation of the effective potential  at the conformal phase transition point: $V^{(4)}(\phi)= -( 4N_F N_C m_F^4/\pi^4) \chi^4 (\ln \chi-1/4)$, in precise agreement with
Eq.(\ref{WTC}) through Eq.(\ref{PCDC2}). See Eq.(65) in Ref.~\cite{Miransky:1996pd}} 
 responsible for the {\it trace anomaly of dimension 4 operator} this time:~\cite{Matsuzaki:2012mk,Matsuzaki:2015sya}
 \begin{eqnarray}
  {\cal L}_{\rm WTC}
 &=&\chi^2 \cdot \left[ \frac{1}{2} \left(\partial_\mu \phi\right)^2  +\frac{F_\pi^2}{4}{\rm tr} \left({\cal D}_\mu U {\cal D}^\mu U^\dagger\right)\right] -V^{(4)}(\phi)-V^{(\rm SM)}(\phi) \nonumber \,,\\
 V^{(4)}(\phi)&=&-\ln \chi\cdot \frac{\beta^{(\rm NP)}(\alpha)}{4\alpha} G_{\mu\nu}^2= \frac{M_\phi^2 F_\phi^2}{4} \chi^4 \left(\ln\chi-\frac{1}{4}\right)\nonumber\\
V^{(\rm SM)}(\phi)&=& - \chi^{2-\gamma_m} \left(m_f \chi \bar f f\right)  -\ln \chi \left[\frac{\beta_F(\alpha_s)}{4\alpha_s} {G^{({\rm gluon})}_{\mu\nu}}^2 
+ \frac{\beta_F(\alpha_e)}{4\alpha_e} {F^{({\gamma})}_{\mu\nu}}^2\right]\,,\,\nonumber\\
 \chi &= &\exp \left(\frac{\phi}{F_\phi}\right)\,,
 \label{WTC}
 \end{eqnarray} 
where {\it $F_\phi\ne F_\pi=v/\sqrt{N_D}=v/\sqrt{N_F/2}$ in general} in contrast to the SM Higgs case $F_\phi=F_\pi=v$, the electroweak gauging was done as usual  $\partial_\mu U\Rightarrow {\cal D}_\mu U= \partial_\mu U -i g_2 W_\mu U +i g_1 U B_\mu$, and we have
added  $V^{(\rm SM)}(\phi)$, the scale symmetry breaking terms related  to the SM particles arising from the technifermion contributions: mass term of the SM fermion $f$,  (one loop)  technifermion  contributions to the 
trace anomaly for  the gluon and photon operators, with  $\beta_F(g_s) = \frac{g_s^3}{(4\pi)^2} \frac{4}{3} N_C$ and  $\beta_F(e) = \frac{e^3}{(4\pi)^2} \frac{16}{9} N_C$. 
It is obvious that ${\theta_\mu^\mu}^{({\rm TC})}=- \delta_D V^{(4)}(\phi) = \beta^{(\rm NP)}(\alpha)/(4\alpha)\cdot G_{\mu\nu}^2=- (M_\phi^2F_\phi^2/4) \chi^4$ up to total derivative, corresponding to
the PCDC with $d_\theta=4$ ($\langle \chi\rangle=1$), Eq.(\ref{PCDC2}). 

The technidilaton potential   $V^{(4)}(\phi)$ is expanded in $\phi/F_\phi$:
\begin{equation} 
V^{(4)}(\phi)
= - \frac{M_\phi^2 F_\phi^2}{16} +\frac{1}{2}M_\phi^2\,\phi^2 +\frac{4}{3} \frac{M_\phi^2}{F_\phi} \,\phi^3 
+ 2 \frac{M_\phi^2}{F_\phi^2}\, \phi^4 
+ \cdots 
\,,
\label{dilatonpotential}
\end{equation}
which shows a remarkable fact that in the anti-Veneziano limit 
the technidilaton  self couplings (trilinear and quartic couplings) are highly suppressed:
\begin{equation}
\lambda_{\phi^3}=4M_\phi^2/(3F_\phi) \sim 1/\sqrt{N_F N_C}\rightarrow 0\,, \quad
\lambda_{\phi^4}=2 M_\phi^2/F_\phi^2 \sim 1/(N_F N_C)\rightarrow 0\,,
\label{weakcouplingTD}
\end{equation} 
by $M_\phi/F_\phi \sim 1/\sqrt{N_F N_C}$ and $M_\phi \sim N_F^0 N_C^0$. 
Numerically, we may compare the technidilaton self couplings
with those
of the SM Higgs with $m_h=M_\phi=125$ GeV for $v/F_\phi \simeq 1/5$ in the one-family model ($N_F=8, N_C=4)$: \cite{Matsuzaki:2015sya}
\begin{eqnarray} 
\frac{\lambda_{\phi^3}}{\lambda_{h_{\rm SM}^3}}\Bigg|_{M_\phi=m_h} 
&=& \frac{\frac{4 M_\phi^2}{3 F_\phi}}{\frac{m_h^2}{2 v}} \Bigg|_{M_\phi=m_h}
\simeq \frac{8}{3} \left( \frac{v}{F_\phi}\right) \simeq 0.5 \,, \nonumber \\ 
\frac{\lambda_{\phi^4}}{\lambda_{h_{\rm SM}^4}} \Bigg|_{M_\phi = m_h} 
&=& \frac{\frac{2 M_\phi^2}{F_\phi^2}}{\frac{m_h^2}{8 v^2}}\Bigg|_{M_\phi=m_h} 
= 16 \left( \frac{v}{F_\phi} \right)^2 \simeq 0.6 
\,. 
\label{selfcouplings:0}
\end{eqnarray}
This shows that the {\it technidilaton self couplings, although generated by the strongly coupled interactions, are even smaller than those of the 
SM Higgs}, a salient feature of the approximate scale symmetry in the ant-Veneziano limit !!

The coupling of the technidilaton ($M_\phi=125$ GeV) to the SM particles
can be seen by expanding $\chi= 1+\phi/F_\phi +(1/2!)(\phi/F_\phi)^2+\cdots$ in Eq.(\ref{WTC}): 
\begin{eqnarray} 
  \frac{g_{\phi WW/ZZ}}{g_{ h_{\rm SM} WW/ZZ }}   
 =
 \frac{g_{\phi ff}}{g_{h_{\rm SM} ff}}  \,
 = 
  \frac{v}{F_\phi} 
  \,.  
  \label{WWZZ}
\end{eqnarray} 
 \begin{eqnarray} 
\frac{g_{\phi gg}}{g_{h_{\rm SM} gg}} 
\simeq 
\frac{v}{F_\phi} 
\cdot 
\left( 1 + 2 N_C \right),\,
\frac{g_{\phi \gamma\gamma}}{g_{h_{\rm SM} \gamma\gamma}} 
\simeq \frac{v}{F_\phi} 
\cdot 
 \left( \frac{63 -  16}{47} - \frac{32}{47} N_C \right)  
\,,  \label{g-dip-dig}
\end{eqnarray} 
where besides the technifermions loop, only the top and W of the SM contributions  were included at one-loop.
Note the couplings in Eq.(\ref{WWZZ}) with $v/F_\phi \sim1/5$ are even weaker than the SM Higgs, which are however compensated by
those in Eq.(\ref{g-dip-dig}) for  $gg$ and $\gamma\gamma$ rather enhanced
by the  extra  loop contributions of the technifermions other than the SM particles, particularly for large $N_C$, resulting in signal strength similar to the SM Higgs within the current experimental accuracy.  

In fact the current LHC data for 125 GeV Higgs are fit by the technidilaton as good as by the SM Higgs,
particularly for $N_F=8, N_C=4$, i.e.,  near the anti-Veneziano limit.~\cite{Matsuzaki:2012mk}  
Most recent detailed analyses are given in Ref. \cite{Matsuzaki:2015sya}.
It should be mentioned here that the one-family model will be most naturally imbedded into the ETC in the case for $N_C=4$ \cite{Kurachi:2015bva}.
More precise data at LHC Run II will discriminate among them, SM Higgs or technidilaton. 

What about  the technipions? In the walking technicolor with $N_D =N_F/2>1$, the spontaneous breaking of the chiral symmetry larger than $SU(2)_L\times SU(2)_R$ produces NG bosons (technipions) more than 3 to
be absorbed into W/Z. Let us take the one-family model with $N_F=8$, which has colored techniquarks (3 weak doublets) $Q^a_i$ and non-colored technileptons (one weak-doublet) $L_i$
($a=1,2,3;i=1,2$), the resultant chiral symmetry being $SU(8)_L \times SU(8)_R$ \cite{Dimopoulos:1979sp}. There are 63 technipions, 60 of which are  unabsorbed  technipions.   All of them acquire the mass from the explicit chiral symmetry breaking due to the SM gauge and ETC gauge interactions. Due to the large anomalous dimension $\gamma_m\simeq 1$, the mass of them are all
enhanced to TeV region \cite{Kurachi:2014xla}, which will be discovered at LHC Run II.
(After this symposium, 750 GeV diphoton events were reported at LHC \cite{750}, which can be identified with  the color-singlet and iso-singlet (not flavor-singlet)  technipion
$P^0$ \cite{Matsuzaki:2015che,Lebiedowicz:2016lmn}.
 If it is the case, $M_{P^0}=750$ GeV, the model predicts another nearby color-singlet technipion $P^i$ (iso-triplet one), with mass $M_{P^i}= \sqrt{\frac{8}{5}} M_{P^0} \simeq
950$ GeV, a salient prediction of the one-family model independently of the dynamical details \cite{Kurachi:2014xla}.) 

Another signature of the walking technicolor is the prediction of higher resonances such as the spin 1 boson, the walking techni-$\rho$, 
walking techni-$a_1$, etc..
The straightforward $N_F$ extension of Eq.(\ref{SHLS})  is also obvious:  Eq.(\ref{WTC}) is gauge equivalent to the {\it scale-invariant HLS Lagrangian} explicitly constructed for one-family walking technicolor with $N_F=8$ \cite{Kurachi:2014qma}: 
\begin{eqnarray}
{\cal L}_{\rm s-HLS} &=& {\cal L}_{\rm WTC}+ {\cal L}_{\rm Kinetic} \left(V_\mu\right)\,,\nonumber \\
{\cal L}_{\rm WTC}&=& \chi^2 \cdot \left(\frac{1}{2} \left(\partial_\mu \phi\right)^2 + {\cal L}_A+ a {\cal L}_V\right) -V^{(4)}(\phi) \,,
\label{SHLS2}
\end{eqnarray} 
where the HLS gauge bosons $V_\mu$ in the mass term $ a \chi^2 {\cal L}_V=a e^{2\phi/F_\phi} {\rm tr} (g_{_{\rm HLS}} V_\mu +\cdots)^2$ are the bound states of the walking technicolor, the walking techni-$\rho$,
with {\it mass term $M_V^2=a g_{_{\rm HLS}}^2 F_\pi^2$ being scale-invariant} thanks to the overall technidilaton factor $\chi^2$, as mentioned before. 
The loop expansion is formulated as the scale-invariant HLS perturbation theory \cite{Fukano:2015zua} in the same way as the scale-invariant chiral perturbation theory \cite{Matsuzaki:2013eva}, a straightforward extension of the (non-scale-invariant) HLS perturbation theory \cite{Harada:2003jx}.

In fact there has been reported an interesting 2 TeV diboson event at LHC \cite{Aad:2015owa}. 
We have shown \cite{Fukano:2015hga}
 that this would be 
the most natural candidate for the Drell-Yan produced walking techni-$\rho$, the color-singlet iso-triplet $\rho^{i}$,  as a gauge boson of the HLS described 
by the scale-invariant HLS model in Eq.(\ref{SHLS2}) \cite{Kurachi:2014qma}.  
We further found \cite{
Fukano:2015zua} 
that a salient feature of this possibility is
the scale symmetry which forbids the decay of the walking techni-$\rho$ to the 125 GeV Higgs (technidilaton) plus $W/Z$ (what we called
``conformal barrier'') in the same way as the hidden vector in the SM Higgs, Eq. (\ref{Conformalbarrier}),  in sharp contrast to the popular ``equivalence theorem''.

The HLS is readily extendable to include techni-$a_1$, etc. \cite{Bando:1984ej,Bando:1987br,Harada:2003jx}, with an infinite set of the HLS tower being equivalent to the deconstructed extra dimension \cite{ArkaniHamed:2001ca} and/or the holographic models \cite{Son:2003et}, and the scale-invariant version of them are also straightforward and  {\it mass term of all the higher HLS vector bosons
are scale-invariant}, an outstanding characterization in sharp contrast  to other formulations for the spin 1 bosons.  The conformal barrier  applies not only to the techni-$\rho$ but
also to  all the higher vector/axialvector resonances as the HLS gauge bosons, having scale-invariant mass term. The LHC Run II will tell us whether or not it is the case. 

\section{Walking Technicolor on the Lattice}

Finally, I would like to make a brief review on the lattice studies on the walking technicolor particularly done by our group, the
LatKMI Collaboration \cite{Aoki:2012eq,Aoki:2013xza,Aoki:2013zsa,Aoki:2014oha}.
 Since the dynamics is essentially nonperturbative,
reliable calculations would be eventually done by the lattice simulations.
In fact there has been extensive activity in the lattice simulations of the candidate theories for the walking technicolor, particularly the large $N_F$ QCD, i.e.,
the $SU(N_C)$ gauge theory with $N_F$ degenerate flavors of fundamental representation fermions, eventually extrapolated to the
chiral limit  \cite{Lattice}. Among others particular interest has been paid to the $N_f=8$ and $N_F=12$ for $N_C=3$, partly because the infrared fixed point $\alpha_*$ in the two-loop beta function exists for $N_F \gtrsim 8$ ($N_C=3$),
 and the SSB criticality condition  $\alpha_*=\alpha_{\rm cr}$ for the ladder 
 result $\alpha_{\rm cr}=\pi/(3 C_2)$ is fulfilled for $N_F \simeq 12$ \cite{Appelquist:1996dq}, so that it is expected that the walking theory might exist  somewhere around $8<N_F<12$. Inexpensive simulations are mostly done in the staggered fermion, $N_F=4,8,12,16$, within the asymptotically free
 cases.  

 The LatKMI Collaboration started in 2010 for the lattice simulations on the possible candidate for the walking technicolor by 
systematic studies of the $N_F=16,12,8, 4$ on the same lattice set-up, using the HISQ (Highly Improved Staggered Quarks) 
action with tree-level Symanzik gauge action. We have mainly focused on the low-lying fermionic bound states (plus some gluonic ones), i,e., pseudoscalar (denoted as $\pi$), scalar ($\sigma,a_0$), vector ($\rho$), axialvector mesons ($a_1$), and nucleon-like states ($N, N^*$), etc. and particularly the flavor-singlet scalar $\sigma$ as a candidate for the technidilaton.

We found \cite{Aoki:2012eq} that $N_F=12$ is consistent with the conformal window indicating no spontaneous chiral symmetry breaking, satisfying  the universal hyperscaling relation $M\sim m_f^{1/(1+\gamma_m)}$
for all the observed quantities (with $\gamma_m \sim 0.4\ll 1$), which is in agreement with the results of many other groups \cite{Lattice}.  

We further found \cite{Aoki:2013xza} that $N_F=8$, in comparison with $N_F=12$ and $N_F=4$ (up-dated in \cite{Aoki:2015zny}),
 is consistent with the SSB phase with remnants of the conformality (approximately universal hyperscaling relation {\it except for the NG boson pion mass}) with a large anomalous dimension
\begin{equation}
 \gamma_m \simeq 1\,,
 \end{equation}
  namely the walking theory, which was confirmed by other groups \cite{Appelquist:2014zsa}.

A remarkable result of the LatKMI Collaboration is the {\it discovery of a light flavor-singlet scalar on the lattice}, lighter than the pion 
in  $N_F=12$ \cite{Aoki:2013zsa}.  
This  $N_F=12$ result of us was confirmed by 
other groups \cite{Fodor:2014pqa}. Since the theory is consistent with the conformal window without SSB,  such a light flavor-singlet scalar in $N_F=12$ may have no
direct relevance to the technidilaton as a composite Higgs. Nevertheless, it is suggestive that the conformal dynamics may produce the dilatonic scalar, which was generated only by
the explicit breaking fermion bare mass $m_0$ put on the lattice. 

Furthermore, we made an outstanding discovery that in $N_F=8$ there exists  a light flavor-singlet scalar with mass comparable to the pion \cite{Aoki:2014oha}.
 See Fig. \ref{fig:Nf8scalar}.
 \begin{figure} [h]
 \begin{center}
 % \begin{wrapfigure}{r}{6.6cm}
  \includegraphics[width=6cm]{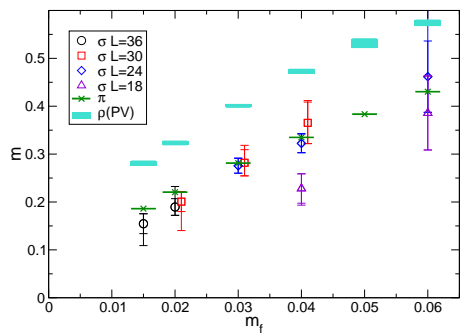}
\caption{Flavor-singlet scalar meson denoted as $\sigma$ in $N_f=8$ QCD HISQ with $\beta=6/g^2=3.8$, in comparison with the NG boson pion $\pi$ and the vector meson $\rho$, for various
values of the degenerate fermion bare mass $m_f$ on the lattice. Lattice volumes are $L^3\times T=36^3\times 48, 30^3\times 40, 24^3\times 32, 18^3\times 24$.  }
\label{fig:Nf8scalar}
\end{center} 
 \end{figure} 
Our $N_F=8$ results were also confirmed by other group \cite{Appelquist:2016viq}.  
Since $N_F=8$ seems to be a walking theory in the SSB phase as mentioned above,  the light flavor-singlet scalar is particularly attractive as a candidate for the technidilaton.  
Also $N_F=8$ is  of phenomenological 
relevance to the LHC data and of direct relevance to the one-family model as the  most natural model building.
Future confirmation of our results is highly desired. Also $N_C=4$ simulations should be studied for various reasons as mentioned before.

\section{Conclusion}
We have discussed the modern version of the Nambu's path to the origin of mass, namely the dynamical symmetry breaking of the electroweak symmetry, which 
may account for the origin of the otherwise mysterious input mass parameter (tachyon mass ) of the SM Higgs Lagrangian. The dynamics changes the vacuum by the strong coupling
attractive forces so as to produce the  mass scale $m_F (\ll \Lambda)$ smaller than the intrinsic scale $\Lambda (\Lambda_{\rm QCD}) $ which is given either at Lagrangian level (like NJL model)
or induced as the trace anomaly when regularizing the quantum theory (like QCD and technicolor). Crucial to the hierarchy $m_F\ll \Lambda$ is the non-zero critical coupling, which yields a large anomalous dimension and infrared conformality even in the
NJL type four-fermion theory having the coupling of explicit mass scale.

We have defined ``strong coupling theories'' as ``those having non-zero critical coupling'' $N_C g_{\rm cr}={\cal O}(1)$, even though its
value could be small $g \sim 1/N_C \ll 1$ in the typical large $N_C$ limit. The NJL model pioneered by Professor Nambu is 
the first and a typical example of such, to be distinguished from its preceding, the BCS theory, which has a zero critical coupling
$g_{\rm cr}=0$.  Existence of such a non-zero critical coupling in gauge theory 
was discovered by Maskawa and Nakajima in the
scale-invariant dynamics, ladder approximation, and became crucial for the walking technicolor with the coupling
$N_C \alpha> N_C \alpha_{\rm cr} ={\cal O}(1)$ in the SSB phase of the scale symmetry as well as the chiral symmetry. 

The existence of the non-zero critical coupling  is actually ``hidden'' even in the QCD which is regarded to have only one phase in the ordinary
situation without signal of the no-zero critical coupling: it manifests itself in  the extreme condition, such as the 
large number of fermions $N_F\gg N_C$ (so as to keep the asymptotic freedom), high temperature, high density, etc.. 

Indeed, it is the large $N_F$ QCD that models the walking technicolor where the large number of fermions give the screening effects and level off of
 the infrared coupling which otherwise blows up due to the gluon anti-screening effects (Caswell-Banks-Zaks infrared fixed point). 
 For large $N_F$ with the fixed point value smaller than the critical coupling,  the SSB phase is gone (what we called conformal  phase transition). 
 Then the infrared scale invariance becomes manifest, dubbed the conformal window. The waking technicolor is close to  just outside of the  conformal
 window. Although $N_f$ and $N_C$ are integers, the anti-Veneziano limit makes the analyses of the phase in the almost continuous parameter $r=N_F/N_C$.

To see the relevance of the infrared conformality, we have argued that the 125 GeV Higgs itself
is a (pseudo-)dilaton, with mass coming from the trace anomaly of dimension 2 due to the Lagrangian parameter,
even
  if it is described by  the Standard Model Higgs Lagrangian (!!). The SM Higgs Lagarangian
    was in fact shown to be equivalent to the scale-invariant nonlinear sigma model with both chiral and scale symmetries 
being nonlinearly realized. The SM Higgs Lagrangian was further shown to be gauge equivalent to the scale-invariant Hidden Local Symmetry (HLS)
Lagrangian which include new massive vector bosons as the gauge bosons of the (spontaneously broken) HLS, 
with the mass term being scale-invariant.  

All these features of the SM Higgs Lagrangian are reminiscent of the conformal UV completion behind the Higgs, the underlying theory with (approximate) scale symmetry with the coupling  so strong as to produce composite states
such as the Higgs (dilaton), new vector bosons (HLS gauge bosons), etc.. We have seen that even the NJL model, though not gauge theory,
 can be regarded as such a
conformal UV completion. 

The walking technicolor, conformal SCGT,  is a gauge theory version of such a typical  
underlying theory, where the 125 GeV Higgs is a composite pseudo-dilaton, technidilaton, with mass coming from the nonperturbative trace anomaly. 
The walking technicolor in the
anti-Veneziano limit  $N_C \rightarrow \infty$ with $N_C \alpha=$fixed $={\cal O}(1)$ and $N_F/N_C=$ fixed ($\gg 1)$ makes
the ladder approximation reasonable, which yields a naturally light and weakly coupled technidilaton through the PCDC:
 \begin{equation}
 M_\phi^2 F_\phi^2=  
-4 \langle \theta_\mu^\mu \rangle 
=-  \frac{\beta(\alpha (\mu^2))}{\alpha(\mu^2)}
\, \langle G_{\mu \nu}^2(\mu^2)\rangle 
\simeq N_C N_F\frac{16 }{\pi^4} m_F^4\,,
\end{equation}
independently of the renormalization point $\mu$,
where the scale symmetry is explicitly broken by the nonperturbative trace anomaly of the dimension 4 operator $G_{\mu\nu}^2$,
which is induced by $m_F$ the 
dynamical mass of the technifermion arising from the simultaneous spontaneous breaking of the scale symmetry and the chiral symmetry.
The technidilaton with mass 125 GeV and coupling by the PCDC relation for the one-family model with $N_F=8, N_c=4$ is consistent with
the present data for the LHC 125 GeV Higgs. 

Lattice results are also encouraging  for the light composite Higgs as the technidilaton in the walking technicolor, particularly for $N_F=8$ QCD,
which correspond to the one-family technicolor as the most straightforward walking technicolor model building.
 
 We will see the fate of the strong coupling theories at LHC Run II, hoping that the Nambu's way will continue forever.

\section*{Acknowledgment}
Main materials given here were developed during the discussions and collaborations with Shinya Matsuzaki. 
I also thank all the members of LatKMI Collaboration for excellent collaborations on the strong coupling gauge theories on the lattice.

%Insert the Acknowledgment text here.


%\bibliographystyle{ptephy}
%\bibliography{sample}
%
% once the .bbl file has been generated then place the text in your article.

\end{document}